\documentclass{article}
\usepackage{amsfonts}
\usepackage{amssymb}
\usepackage{amsmath}

\begin{document}
\author{C. C. Ciob\^{\i}rc\u{a}\thanks{%
e-mail address: ciobarca@central.ucv.ro}, S. O. Saliu\thanks{%
e-mail address: osaliu@central.ucv.ro}, \\
Faculty of Physics, University of Craiova\\
13 A. I. Cuza Str., Craiova 200585, Romania}
\title{Generalized couplings between an Abelian $p$-form and a $\left(
3,1\right) $ mixed symmetry tensor field}
\maketitle

\begin{abstract}
The consistent interactions between a single, free, massless tensor gauge
field with the mixed symmetry $(3,1)$ and an Abelian $p$-form are investigated in the framework of the BRST formalism combined
with cohomological techniques. Under the assumptions on smoothness,
locality, Lorentz covariance, and Poincar\'{e} invariance of the
deformations, supplemented by the requirement that the interacting
Lagrangian is at most second-order derivative, it is proved that for every value $p\geq 1$ of the form degree there are 
consistent couplings between the Abelian form and the massless $(3,1)$ gauge field.
\end{abstract}

\section{Introduction}

Tensor fields in \textquotedblleft exotic" representations of the Lorentz
group, characterized by a mixed Young symmetry type~\cite%
{curt1,curt2,aul,labast1,labast2,burd,zinov1}, held the attention lately on
some important issues, like the dual formulation of field theories of spin
two or higher~\cite%
{dualsp1,dualsp2,dualsp2a,dualsp2b,dualsp3,dualsp4,dualsp5}, the
impossibility of consistent cross-interactions in the dual formulation of
linearized gravity~\cite{lingr}, or a Lagrangian first-order approach~\cite%
{zinov2,zinov3} to some classes of massless or partially massive mixed
symmetry type tensor gauge fields, suggestively resembling to the tetrad
formalism of General Relativity. An important matter related to mixed
symmetry type tensor fields is the study of their consistent interactions,
among themselves as well as with higher-spin gauge theories~\cite%
{high1,high2,high3,high4,7,9,kk,3,4}. The most efficient approach to this
problem is the cohomological one, based on the deformation of the solution
to the master equation~\cite{def}. The purpose of this paper is to
investigate the consistent interactions between a single free massless
tensor gauge field $t_{\lambda \mu \nu |\alpha }$ with the mixed symmetry of
a two-column Young diagram of the type $(3,1)$ and one Abelian form $A_{\mu
_{1}...\mu _{p}}$. It is worth mentioning the duality of the free massless
tensor gauge field $t_{\lambda \mu \nu |\alpha }$ to the Pauli-Fierz theory
in $D=6$ dimensions and, in this respect, the recent developments concerning
the dual formulations of linearized gravity from the perspective of $M$%
-theory~\cite{mth1,mth2,mth3}. Our analysis relies on the deformation of the
solution to the master equation by means of cohomological techniques with
the help of the local BRST cohomology, whose component in the $(3,1)$ sector
has been reported in detail in~\cite{noijhep31}. Under the hypotheses of
smoothness in the coupling constant, locality, Lorentz covariance, and
Poincar\'{e} invariance of the deformations, combined with the preservation
of the number of derivatives on each field, we prove that for every Abelian $%
p$-form there exists a deformation of the solution to the master equation,
which provides nontrivial cross-couplings. This case corresponds to a $%
\left( p+4\right) $-dimensional space-time and is described by a deformed
solution that stops at order two in the coupling constant. The interacting
Lagrangian action contains only mixing-component terms of order one and two
in the coupling constant. At the level of the gauge transformations, only
those of the Abelian form are modified at order one in the coupling constant
with a term linear in the antisymmetrized first-order derivatives of some
gauge parameters from the $(3,1)$ sector such that the gauge algebra and the
reducibility structure of the coupled model are not modified during the
deformation procedure, being the same like in the case of the starting free
action. It is interesting to note that if we require the PT invariance of
the deformed theory, then no interactions occur. Although it is not possible
to construct interactions that deform the gauge algebra, our results (see 
\cite{31vect}) are interesting since these seem to be the first cases where
mixed symmetry type tensor fields allow nontrivial cross-couplings.

\section{The free theory and the BRST symmetry}

We start from a \textquotedblleft free\textquotedblright\ Lagrangian action\
written as a sum between the Lagrangian action $S_{0}\left[ t_{\mu \nu
\lambda |\rho }\right] $ of the tensor field with the mixed symmetry $\left(
3,1\right) $ and the Lagrangian action $S^{A}\left[ A_{\mu _{1}...\mu _{p}}%
\right] $ of an abelian p-form,%
\begin{equation}
S_{0}\left[ t_{\mu \nu \lambda |\rho },A_{\mu _{1}...\mu _{p}}\right] =S_{0}%
\left[ t_{\mu \nu \lambda |\rho }\right] +S_{0}^{A}\left[ A_{\mu _{1}...\mu
_{p}}\right] ,  \label{tp1}
\end{equation}%
where 
\begin{equation}
S_{0}^{A}\left[ A_{\mu _{1}...\mu _{p}}\right] =-\frac{1}{2\left( p+1\right)
!}\int d^{D}xF_{\mu _{1}...\mu _{p+1}}F^{\mu _{1}...\mu _{p+1}}  \label{tp2}
\end{equation}%
and 
\begin{eqnarray}
S_{0}\left[ t_{\lambda \mu \nu |\alpha }\right] &=&\int d^{D}x\left( \frac{1%
}{2}\left( \left( \partial ^{\rho }t^{\lambda \mu \nu |\alpha }\right)
\left( \partial _{\rho }t_{\lambda \mu \nu |\alpha }\right) -\left( \partial
_{\alpha }t^{\lambda \mu \nu |\alpha }\right) \left( \partial ^{\beta
}t_{\lambda \mu \nu |\beta }\right) \right) \right.  \notag \\
&&-\frac{3}{2}\left( \left( \partial _{\lambda }t^{\lambda \mu \nu |\alpha
}\right) \left( \partial ^{\rho }t_{\rho \mu \nu |\alpha }\right) +\left(
\partial ^{\rho }t^{\lambda \mu }\right) \left( \partial _{\rho }t_{\lambda
\mu }\right) \right)  \notag \\
&&\left. +3\left( \left( \partial _{\alpha }t^{\lambda \mu \nu |\alpha
}\right) \left( \partial _{\lambda }t_{\mu \nu }\right) +\left( \partial
_{\rho }t^{\rho \mu }\right) \left( \partial ^{\lambda }t_{\lambda \mu
}\right) \right) \right) .  \label{t1}
\end{eqnarray}%
The dimension of the space-time satisfies the inequality%
\begin{equation}
D\geq \max \left( 5,p+1\right) .
\end{equation}%
The field strength of the abelian form from the formula (\ref{tp2}) is
defined in the standard manner,%
\begin{equation}
F_{\mu _{1}...\mu _{p+1}}=\partial _{\lbrack \mu _{1}}A_{\mu _{2}...\mu
_{p+1}]},\;D>p.  \label{tp3}
\end{equation}%
\textbf{\ }Everywhere in this paper we employ the flat Minkowski metric of
`mostly plus' signature $\sigma ^{\mu \nu }=\sigma _{\mu \nu }=\left(
-,++++\cdots \right) $.\textbf{\ }We remember $A_{\mu _{1}...\mu _{p}}$ is
an antisymmetric tensor and the mixed symmetry $\left( 3,1\right) $ of the
tensor field $t_{\lambda \mu \nu |\alpha }$ means it is antisymmetric in the
first three indices and satisfies the identity 
\begin{equation}
t_{\left[ \lambda \mu \nu |\alpha \right] }\equiv 0.  \label{t4}
\end{equation}

The functional $S_{0}\left[ t_{\mu \nu \lambda |\rho }\right] $ is invariant
to the known gauge transformations (see \cite{noijhep31}) of the tensor
field $t_{\lambda \mu \nu |\alpha }$,%
\begin{equation}
\delta _{\epsilon ,\chi }t_{\lambda \mu \nu |\alpha }=3\partial _{\alpha
}\epsilon _{\lambda \mu \nu }+\partial _{\left[ \lambda \right. }\epsilon
_{\left. \mu \nu \right] \alpha }+\partial _{\left[ \lambda \right. }\chi
_{\left. \mu \nu \right] |\alpha },  \label{tp4}
\end{equation}%
where $\epsilon _{\lambda \mu \nu }$ is an arbitrary, antisymmetric tensor
field and $\chi _{\mu \nu |\alpha }$ has the mixed symmetry $\left(
2,1\right) $ (it is antisymmetric in the first two indices and satisfy the
identity $\chi _{\left[ \mu \nu |\alpha \right] }\equiv 0$). The gauge
symmetries (\ref{tp4}) are off-shell second stage reducible, because the
right side of the formula (\ref{tp4}) vanishes if we use the replacements 
\begin{eqnarray}
\epsilon _{\mu \nu \alpha } &\rightarrow &\epsilon _{\mu \nu \alpha
}^{\left( \omega ,\psi \right) }=-\frac{1}{2}\partial _{\left[ \mu \right.
}\omega _{\left. \nu \alpha \right] },  \label{t12} \\
\chi _{\mu \nu |\alpha } &\rightarrow &\chi _{\mu \nu |\alpha }^{\left(
\omega ,\psi \right) }=\partial _{\left[ \mu \right. }\psi _{\left. \nu %
\right] \alpha }+2\partial _{\alpha }\omega _{\mu \nu }-\partial _{\left[
\mu \right. }\omega _{\left. \nu \right] \alpha },  \label{t12a}
\end{eqnarray}%
where $\omega _{\mu \nu }$ is an arbitrary, antisymmetric tensor field, $%
\Psi _{\mu \nu }$ an arbitrary, symmetric tensor field, while (\ref{t12})-(%
\ref{t12a}) vanish by the replacements%
\begin{eqnarray}
\omega _{\mu \nu } &\rightarrow &\omega _{\mu \nu }^{\left( \theta \right)
}=\partial _{\left[ \mu \right. }\theta _{\left. \nu \right] },  \label{t15}
\\
\psi _{\mu \nu } &\rightarrow &\psi _{\mu \nu }^{\left( \theta \right)
}=-3\partial _{\left( \mu \right. }\theta _{\left. \nu \right) }.
\label{t15a}
\end{eqnarray}%
The most general object invariant to the gauge transformations (\ref{tp4})
is the curvature tensor, defined by%
\begin{equation}
K^{\lambda \mu \nu \xi |\alpha \beta }=\partial ^{\alpha }\partial ^{\left[
\lambda \right. }t^{\left. \mu \nu \xi \right] |\beta }-\partial ^{\beta
}\partial ^{\left[ \lambda \right. }t^{\left. \mu \nu \xi \right] |\alpha
}=^{\beta }\partial ^{\left[ \lambda \right. }t^{\left. \mu \nu \xi \right] |%
\left[ \beta ,\alpha \right] },  \label{t20}
\end{equation}%
which has the mixed symmetry $\left( 4,2\right) $ and satisfies Bianchi type
identities: algebraic, $K^{\left[ \lambda \mu \nu \xi |\alpha \right] \beta
}\equiv 0$, and differential, $\partial ^{\left[ \kappa \right. }K^{\left.
\lambda \mu \nu \xi \right] |\alpha \beta }\equiv 0,\;K^{\lambda \mu \nu \xi
|\left[ \alpha \beta ,\gamma \right] }\equiv 0$. We can express the field
equations of $t_{\lambda \mu \nu |\rho }$ with the help of the curvature
tensor, 
\begin{equation}
\frac{\delta S_{0}}{\delta t_{\lambda \mu \nu |\alpha }}\equiv -T^{\lambda
\mu \nu |\alpha }\approx 0,  \label{t16}
\end{equation}%
\begin{equation}
T^{\lambda \mu \nu |\alpha }=K^{\lambda \mu \nu |\alpha }-\frac{1}{2}\sigma
^{\alpha \left[ \lambda \right. }K^{\left. \mu \nu \right] }.  \label{t26}
\end{equation}%
An important property of field equations (\ref{t16}) is they can be written
as 
\begin{equation}
T^{\lambda \mu \nu |\alpha }=\partial _{\xi }\partial _{\beta }\Phi
^{\lambda \mu \nu \xi |\alpha \beta },  \label{t32f}
\end{equation}%
where $\Phi ^{\lambda \mu \nu \xi |\alpha \beta }$ is an antisymmetric
tensor, separately, in the first two indices, respectively in the last two, 
\begin{eqnarray}
\Phi ^{\lambda \mu \nu \xi |\alpha \beta } &=&-\sigma ^{\alpha \left[
\lambda \right. }\sigma ^{\left. \mu \right] \beta }t^{\nu \xi }-\sigma
^{\alpha \left[ \nu \right. }\sigma ^{\left. \xi \right] \beta }t^{\lambda
\mu }+\sigma ^{\alpha \left[ \lambda \right. }\sigma ^{\left. \nu \right]
\beta }t^{\mu \xi }+\sigma ^{\alpha \left[ \mu \right. }\sigma ^{\left. \xi %
\right] \beta }t^{\lambda \nu }  \notag \\
&&-\sigma ^{\alpha \left[ \lambda \right. }\sigma ^{\left. \xi \right] \beta
}t^{\mu \nu }-\sigma ^{\alpha \left[ \mu \right. }\sigma ^{\left. \nu \right]
\beta }t^{\lambda \xi }+\sigma ^{\alpha \left[ \lambda \right. }t^{\left.
\mu \nu \xi \right] |\beta }-\sigma ^{\beta \left[ \lambda \right.
}t^{\left. \mu \nu \xi \right] |\alpha }.  \label{t32g}
\end{eqnarray}

The functional $S_{0}^{A}\left[ A_{\mu _{1}...\mu _{p}}\right] $ has the
gauge symmetries%
\begin{equation}
\delta _{\overset{\left( 1\right) }{\rho }}A_{\mu _{1}...\mu _{p}}=\partial
_{\left[ \mu _{1}\right. }\overset{\left( 1\right) }{\rho }_{\left. \mu
_{2}...\mu _{p}\right] },  \label{tp5}
\end{equation}%
where the gauge parameter $\overset{\left( 1\right) }{\rho }_{\mu _{2}...\mu
_{p}}$ is an arbitrary, antisymmetric tensor field. The gauge
transformations (\ref{tp5}) are off-shell reducible, vanishing if we use the
replacements%
\begin{equation}
\overset{\left( 1\right) }{\rho }_{\mu _{1}...\mu _{p-1}}=\partial _{\left[
\mu _{1}\right. }\overset{\left( 2\right) }{\rho }_{\left. \mu _{2}...\mu
_{p-1}\right] }\Rightarrow \delta _{\overset{\left( 1\right) }{\rho }\left( 
\overset{\left( 2\right) }{\rho }\right) }A_{\mu _{1}...\mu _{p}}\equiv 0,
\label{tp6}
\end{equation}%
where $\overset{\left( 2\right) }{\rho }_{\mu _{2}...\mu _{p-1}}$ is an
arbitrary, antisymmetric tensor field. The relation (\ref{tp6}) represents
the first order reducibility, the Lagrangian action $S^{A}\left[ A_{\mu
_{1}...\mu _{p}}\right] $ describing a $p-1$ reducible gauge theory. The
reducibility relation in the $k-1$ order is%
\begin{equation}
\overset{\left( k\right) }{\rho }_{\mu _{1}\ldots \mu _{p-k}}=\partial _{%
\left[ \mu _{1}\right. }\overset{\left( k+1\right) }{\rho }_{\left. \mu
_{1}\ldots \mu _{p-k}\right] }\Rightarrow \overset{\left( k-1\right) }{\rho }%
\left( \overset{\left( k\right) }{\rho }\right) \equiv 0,\ k=\overline{2,p}%
\text{.}  \label{tp7}
\end{equation}%
(For $k=p$, $\overset{\left( p\right) }{\rho }$ has no indices.) It follows
the theory described by $S_{0}^{A}\left[ A_{\mu _{1}...\mu _{p}}\right] $
has the Cauchy order $p+1$, so the \textquotedblleft free\textquotedblright\
theory\ described by the action (\ref{tp1}) has the Cauchy order $p+1$ if $%
p>3$, or $4$ if $p\leq 3$. This affirmation will be important later, when we
will need the cohomology $H_{I}\left( \delta |d\right) $. The field strength 
$F_{\mu _{1}...\mu _{p+1}}$ defined in (\ref{tp3}) represents for the
abelian p-form the analogous of the curvature tensor for the tensor field $%
t_{\lambda \mu \nu |\alpha }$, being the most general object invariant to
the gauge transformations (\ref{tp5}). It can be used to construct the
action (\ref{tp2}) and the field equations for the abelian p-form,%
\begin{equation}
\frac{\delta S_{0}^{A}}{\delta A^{\mu _{1}...\mu _{p}}}=\frac{1}{p!}\partial
^{\lambda }F_{\lambda \mu _{1}...\mu _{p}}.  \label{tp7b}
\end{equation}

The BRST complex corresponding to the tensor field $t_{\lambda \mu \nu
|\alpha }$\textbf{\ }contains the fermionic ghosts $\left\{ \eta _{\lambda
\mu \nu },\mathcal{G}_{\mu \nu |\alpha }\right\} $ asociated to the gauge
parameters $\left\{ \epsilon _{\lambda \mu \nu },\chi _{\mu \nu |\alpha
}\right\} $ from (\ref{tp4}), the bosonic ghosts $\left\{ C_{\mu \nu },%
\mathcal{C}_{\nu \alpha }\right\} $ due to the reducibility parameters in
order one $\left\{ \omega _{\mu \nu },\psi _{\nu \alpha }\right\} $ from (%
\ref{t12})-(\ref{t12a}), the fermionic ghosts $C_{\nu }$ corresponding to the
reducibility parameters in order two $\theta _{\nu }$ from (\ref{t15})-(\ref%
{t15a}), the antifields $t^{\ast \lambda \mu \nu |\alpha }$ of the tensor
field $t_{\lambda \mu \nu |\alpha }$ and the antifields $\left\{ \eta ^{\ast
\lambda \mu \nu },\mathcal{G}^{\ast \mu \nu |\alpha }\right\} $, $\left\{
C^{\ast \mu \nu },\mathcal{C}^{\ast \nu \alpha }\right\} $, $C^{\ast \nu }$
associated to the ghosts. The ghosts have the same properties as the
associated reducibility parameters, 
\begin{eqnarray}
\eta _{\lambda \mu \nu } &=&-\eta _{\mu \lambda \nu }=-\eta _{\lambda \nu
\mu }=-\eta _{\nu \mu \lambda },\;\mathcal{G}_{\mu \nu |\alpha }=-\mathcal{G}%
_{\nu \mu |\alpha },\;\mathcal{G}_{\left[ \mu \nu |\alpha \right] }\equiv 0,
\label{t33} \\
C_{\mu \nu } &=&-C_{\nu \mu },\;\mathcal{C}_{\nu \alpha }=\mathcal{C}%
_{\alpha \nu },  \label{t34}
\end{eqnarray}%
and, also, the antifields have the properties of the fields and the ghosts
associated to, 
\begin{equation}
t^{\ast \lambda \mu \nu |\alpha }=-t^{\ast \mu \lambda \nu |\alpha
}=-t^{\ast \lambda \nu \mu |\alpha }=-t^{\ast \nu \mu \lambda |\alpha
},\;t^{\ast \left[ \lambda \mu \nu |\alpha \right] }=0,  \label{t35}
\end{equation}%
\begin{eqnarray}
\eta ^{\ast \lambda \mu \nu } &=&-\eta ^{\ast \mu \lambda \nu }=-\eta ^{\ast
\lambda \nu \mu }=-\eta ^{\ast \nu \mu \lambda },\;\mathcal{G}^{\ast \mu \nu
|\alpha }=-\mathcal{G}^{\ast \nu \mu |\alpha },\;\mathcal{G}^{\ast \left[
\mu \nu |\alpha \right] }\equiv 0,  \label{t36} \\
C^{\ast \mu \nu } &=&-C^{\ast \nu \mu },\;\mathcal{C}^{\ast \nu \alpha }=%
\mathcal{C}^{\ast \alpha \nu }.  \label{t37}
\end{eqnarray}%
The BRST generators of the complex corresponding to the abelian p-form are
the $p$-form $A_{\mu _{1}...\mu _{p}}$ and its antifield $A_{\mu _{1}...\mu
_{p}}^{\ast }$, the ghosts $\left( \overset{\left( k\right) }{\xi }_{\mu
_{1}...\mu _{p-k}}\right) _{k=\overline{1,p}}$ related to the gauge
parameters and the reducibility functions from (\ref{tp5})-(\ref{tp7}), the
antifields $\left( \overset{\left( k\right) }{\xi }_{\mu _{1}...\mu
_{p-k}}^{\ast }\right) _{k=\overline{1,p}}$ associated to the ghosts (all
these generators are antisymmetric tensors). The antifields have the
properties%
\begin{eqnarray}
\varepsilon \left( A_{\mu _{1}...\mu _{p}}^{\ast }\right)  &=&1,\
\varepsilon \left( \overset{\left( k\right) }{\xi }_{\mu _{1}...\mu
_{p-k}}^{\ast }\right) =\left( k+1\right) \mathrm{\bmod}2,  \label{tp8}
\\
\mathrm{agh}\left( A_{\mu _{1}...\mu _{p}}^{\ast }\right)  &=&1,\ \mathrm{pgh%
}\left( A_{\mu _{1}...\mu _{p}}^{\ast }\right) =0,  \label{tp8b} \\
\mathrm{agh}\left( \overset{\left( k\right) }{\xi }_{\mu _{1}...\mu
_{p-k}}^{\ast }\right)  &=&k+1,\ \mathrm{pgh}\left( \overset{\left( k\right) 
}{\xi }_{\mu _{1}...\mu _{p-k}}^{\ast }\right) =0.  \label{tp8c}
\end{eqnarray}%
(We remember: $\varepsilon =$ Grassmann parity, $\mathrm{agh}=$ antighost
number, $\mathrm{pgh}=$ pureghost number.) The ghosts have the properties%
\begin{equation}
\varepsilon \left( \overset{\left( k\right) }{\xi }_{\mu _{1}...\mu
_{p-k}}\right) =k \bmod 2,\ \mathrm{pgh}\left( \overset{\left( k\right) }{%
\xi }_{\mu _{1}...\mu _{p-k}}\right) =k,\ \mathrm{agh}\left( \overset{\left(
k\right) }{\xi }_{\mu _{1}...\mu _{p-k}}\right) =0.  \label{tp9}
\end{equation}%
We know the BRST differential decompose as $s=\delta +\gamma $. The action
of the Koszul-Tate differential $\delta $ on the BRST complex of the abelian
form is given by the formulas 
\begin{equation}
\delta \left( A_{\mu _{1}...\mu _{p}},\,\overset{\left( k\right) }{\xi }%
_{\mu _{1}...\mu _{p-k}}\right) =0,  \label{tp10}
\end{equation}%
\begin{equation}
\delta A_{\mu _{1}...\mu _{p}}^{\ast }=-\frac{1}{p!}\partial ^{\lambda
}F_{\lambda \mu _{1}...\mu _{p}},  \label{tp11}
\end{equation}%
\begin{eqnarray}
\delta \overset{\left( 1\right) }{\xi }_{\mu \mu _{1}...\mu _{p-1}}^{\ast }
&=&-p\partial ^{\mu }A_{\mu \mu _{1}...\mu _{p-1}}^{\ast },  \label{tp12} \\
\delta \overset{\left( k+1\right) }{\xi }_{\mu _{1}...\mu _{p-k-1}}^{\ast }
&=&\left( -\right) ^{k+1}\left( p-k\right) \partial ^{\mu }\overset{\left(
k\right) }{\xi }_{\mu \mu _{1}...\mu _{p-k-1}}^{\ast },\ k=\overline{1,p-1}.
\label{tp12b}
\end{eqnarray}%
while for the exterior longitudinal derivative $\gamma $ one has%
\begin{equation}
\gamma \left( A_{\mu _{1}...\mu _{p}}^{\ast },\,\overset{\left( k\right) }{%
\xi }_{\mu _{1}...\mu _{p-k}}^{\ast }\right) =0,  \label{tp13}
\end{equation}%
\begin{eqnarray}
\gamma A_{\mu _{1}...\mu _{p}} &=&\partial _{\lbrack \mu _{1}}\overset{%
\left( 1\right) }{\xi }_{\mu _{2}...\mu _{p}]},  \label{tp14} \\
\,\gamma \overset{\left( k\right) }{\xi }_{\mu _{1}...\mu _{p-k}}
&=&\partial _{\lbrack \mu _{1}}\overset{\left( k+1\right) }{\xi }_{\mu
_{2}...\mu _{p-k}]},\ k=\overline{1,p-1},  \label{tp14a} \\
\gamma \overset{\left( p\right) }{\xi } &=&0.  \label{tp14b}
\end{eqnarray}%
For the BRST complex of the tensor field $t_{\lambda \mu \nu |\rho }$ one
has the degrees 
\begin{eqnarray}
\mathrm{pgh}\left( t_{\lambda \mu \nu |\alpha }\right)  &=&0,\;\mathrm{pgh}%
\left( \eta _{\lambda \mu \nu }\right) =1=\mathrm{pgh}\left( \mathcal{G}%
_{\mu \nu |\alpha }\right) ,  \label{t40} \\
\mathrm{pgh}\left( C_{\mu \nu }\right)  &=&2=\mathrm{pgh}\left( \mathcal{C}%
_{\nu \alpha }\right) ,\;\mathrm{pgh}\left( C_{\nu }\right) =3,  \label{t41}
\\
\mathrm{pgh}\left( t^{\ast \lambda \mu \nu |\alpha }\right)  &=&\mathrm{pgh}%
\left( \eta ^{\ast \lambda \mu \nu }\right) =\mathrm{pgh}\left( \mathcal{G}%
^{\ast \mu \nu |\alpha }\right) =0,  \label{t42} \\
\mathrm{pgh}\left( C^{\ast \mu \nu }\right)  &=&\mathrm{pgh}\left( \mathcal{C%
}^{\ast \nu \alpha }\right) =\mathrm{pgh}\left( C^{\ast \nu }\right) =0,
\label{t43} \\
\mathrm{agh}\left( t_{\lambda \mu \nu |\alpha }\right)  &=&\mathrm{agh}%
\left( \eta _{\lambda \mu \nu }\right) =\mathrm{agh}\left( \mathcal{G}_{\mu
\nu |\alpha }\right) =0,  \label{t44} \\
\mathrm{agh}\left( C_{\mu \nu }\right)  &=&\mathrm{agh}\left( \mathcal{C}%
_{\nu \alpha }\right) =\mathrm{agh}\left( C_{\nu }\right) =0,  \label{t45} \\
\mathrm{agh}\left( t^{\ast \lambda \mu \nu |\alpha }\right)  &=&1,\;\mathrm{%
agh}\left( \eta ^{\ast \lambda \mu \nu }\right) =2=\mathrm{agh}\left( 
\mathcal{G}^{\ast \mu \nu |\alpha }\right) ,  \label{t46} \\
\mathrm{agh}\left( C^{\ast \mu \nu }\right)  &=&3=\mathrm{agh}\left( 
\mathcal{C}^{\ast \nu \alpha }\right) ,\;\mathrm{agh}\left( C^{\ast \nu
}\right) =4.  \label{t47}
\end{eqnarray}%
and the actions of $\delta $ and $\gamma $ 
\begin{eqnarray}
\gamma t_{\lambda \mu \nu |\alpha } &=&3\partial _{\alpha }\eta _{\lambda
\mu \nu }+\partial _{\left[ \lambda \right. }\eta _{\left. \mu \nu \right]
\alpha }+\partial _{\left[ \lambda \right. }\mathcal{G}_{\left. \mu \nu %
\right] |\alpha }  \notag \\
&=&4\partial _{\alpha }\eta _{\lambda \mu \nu }+\partial _{\left[ \lambda
\right. }\eta _{\left. \mu \nu \alpha \right] }+\partial _{\left[ \lambda
\right. }\mathcal{G}_{\left. \mu \nu \right] |\alpha },  \label{t49}
\end{eqnarray}%
\begin{equation}
\gamma \eta _{\lambda \mu \nu }=-\frac{1}{2}\partial _{\left[ \lambda
\right. }C_{\left. \mu \nu \right] },  \label{t50}
\end{equation}%
\begin{eqnarray}
\gamma \mathcal{G}_{\mu \nu |\alpha } &=&2\partial _{\alpha }C_{\mu \nu
}-\partial _{\left[ \mu \right. }C_{\left. \nu \right] \alpha }+\partial _{%
\left[ \mu \right. }\mathcal{C}_{\left. \nu \right] \alpha }  \notag \\
&=&2\partial _{\left[ \mu \right. }C_{\left. \nu \alpha \right] }-3\partial
_{\left[ \mu \right. }C_{\left. \nu \right] \alpha }+\partial _{\left[ \mu
\right. }\mathcal{C}_{\left. \nu \right] \alpha },  \label{t51}
\end{eqnarray}%
\begin{equation}
\gamma C_{\mu \nu }=\partial _{\left[ \mu \right. }C_{\left. \nu \right]
},\;\gamma \mathcal{C}_{\nu \alpha }=-3\partial _{\left( \nu \right.
}C_{\left. \alpha \right) },\;\gamma C_{\nu }=0,  \label{t52}
\end{equation}%
\begin{equation}
\gamma t^{\ast \lambda \mu \nu |\alpha }=\gamma \eta ^{\ast \lambda \mu \nu
}=\gamma \mathcal{G}^{\ast \mu \nu |\alpha }=\gamma C^{\ast \mu \nu }=\gamma 
\mathcal{C}^{\ast \nu \alpha }=\gamma C^{\ast \nu }=0,  \label{t53}
\end{equation}%
\begin{equation}
\delta t_{\lambda \mu \nu |\alpha }=\delta \eta _{\lambda \mu \nu }=\delta 
\mathcal{G}_{\mu \nu |\alpha }=\delta C_{\mu \nu }=\delta \mathcal{C}_{\nu
\alpha }=\delta C_{\nu }=0,  \label{t54}
\end{equation}%
\begin{equation}
\delta t^{\ast \lambda \mu \nu |\alpha }=T^{\lambda \mu \nu |\alpha
},\;\delta \eta ^{\ast \lambda \mu \nu }=-4\partial _{\alpha }t^{\ast
\lambda \mu \nu |\alpha },  \label{t55}
\end{equation}%
\begin{equation}
\delta \mathcal{G}^{\ast \mu \nu |\alpha }=-\partial _{\lambda }\left(
3t^{\ast \lambda \mu \nu |\alpha }-t^{\ast \mu \nu \alpha |\lambda }\right) ,
\label{t56}
\end{equation}%
\begin{equation}
\delta C^{\ast \mu \nu }=3\partial _{\lambda }\left( \mathcal{G}^{\ast \mu
\nu |\lambda }-\frac{1}{2}\eta ^{\ast \lambda \mu \nu }\right) ,\;\delta 
\mathcal{C}^{\ast \nu \alpha }=\partial _{\mu }\mathcal{G}^{\ast \mu \left(
\nu |\alpha \right) },  \label{t57}
\end{equation}%
\begin{equation}
\delta C^{\ast \nu }=6\partial _{\mu }\left( \mathcal{C}^{\ast \mu \nu }-%
\frac{1}{3}C^{\ast \mu \nu }\right) ,  \label{t58}
\end{equation}%
We know the BRST differential has a canonical action, $s\cdot =\left( \cdot
,S\right) $, generated by the solution of the master equation $\left(
S,S\right) =0$,%
\begin{equation}
S=S^{t}+S^{A},  \label{tp15}
\end{equation}%
where 
\begin{eqnarray}
S^{t} &=&S_{0}\left[ t_{\lambda \mu \nu |\alpha }\right] +\int d^{D}x\left(
t^{\ast \lambda \mu \nu |\alpha }\left( 3\partial _{\alpha }\eta _{\lambda
\mu \nu }+\partial _{\left[ \lambda \right. }\eta _{\left. \mu \nu \right]
\alpha }+\partial _{\left[ \lambda \right. }\mathcal{G}_{\left. \mu \nu %
\right] |\alpha }\right) \right.   \notag \\
&&-\frac{1}{2}\eta ^{\ast \lambda \mu \nu }\partial _{\left[ \lambda \right.
}C_{\left. \mu \nu \right] }+\mathcal{G}^{\ast \mu \nu |\alpha }\left(
2\partial _{\alpha }C_{\mu \nu }-\partial _{\left[ \mu \right. }C_{\left.
\nu \right] \alpha }+\partial _{\left[ \mu \right. }\mathcal{C}_{\left. \nu %
\right] \alpha }\right)   \notag \\
&&\left. +C^{\ast \mu \nu }\partial _{\left[ \mu \right. }C_{\left. \nu %
\right] }-3\mathcal{C}^{\ast \nu \alpha }\partial _{\left( \nu \right.
}C_{\left. \alpha \right) }\right) .  \label{t60}
\end{eqnarray}%
and%
\begin{eqnarray}
S^{A} &=&S_{0}^{A}\left[ A_{\mu _{1}...\mu _{p}}\right] +\int d^{D}x\left(
A^{\ast \mu _{1}\ldots \mu _{p}}\partial _{\left[ \mu _{1}\right. }\overset{%
\left( 1\right) }{\xi }_{\left. \mu _{2}\ldots \mu _{p}\right] }\right.  
\notag \\
&&\left. +\sum_{k=1}^{p-1}\overset{\left( k\right) }{\xi }^{\ast \mu
_{1}\ldots \mu _{p-k}}\partial _{\left[ \mu _{1}\right. }\overset{\left(
k+1\right) }{\xi }_{\left. \mu _{2}\ldots \mu _{p-k}\right] }\right) .
\label{tp16}
\end{eqnarray}

\section{$H\left( \protect\gamma \right) $ and $H\left( \protect\delta %
|d\right) $}

The cohomological method for the computation of the interactions is known
(see \cite{def,noijhep31}) and it is based on the deformation of the
solution (\ref{tp15}) of the master equation. The deformation has to satisfy
also the master equation, so the components of the deformation $\bar{S}%
=S+gS_{1}+g^{2}S_{2}+\ldots $ have to satisfy a chain of equations
\begin{gather}
\left( S_{1},S\right) =0,  \label{tpn01} \\
\frac{1}{2}\left( S_{1},S_{1}\right) +\left( S_{2},S\right) =0,
\label{tpn02} \\
\left( S_{1},S_{2}\right) +\left( S,S_{3}\right) =0,  \label{tpn03} \\
....  \label{tpn04}
\end{gather}%
obtained by the projection of the
equation $\left( \bar{S},\bar{S}\right) =0$ on the different orders of the
coupling constant $g$. The nonintegrated density of the first order
deformation in the coupling constant, $S_{1}=\int d^{D}x\ a$, satisfies the
local equation%
\begin{equation}
sa=\partial _{\mu }j^{\mu }  \label{tpn1}
\end{equation}%
and it has three independent components, 
\begin{equation}
a=a^{\mathrm{t-t}}+a^{\mathrm{p-p}}+a^{\mathrm{t-p}}.  \label{tp17}
\end{equation}%
$a^{\mathrm{t-t}}$, $a^{\mathrm{p-p}}$ and $a^{\mathrm{t-p}}$ are independent
solutions of the equation (\ref{tpn1}). $a^{\mathrm{t-t}}$ is formed only by
objects from the sector of the tensor field $t_{\lambda \mu \nu |\alpha }$, $%
a^{\mathrm{p-p}}$ has components only from the sector of the abelian p-form
and $a^{\mathrm{t-p}}$ should generate the interactions between the tensor
field $t_{\lambda \mu \nu |\alpha }$ and the $p$-form, so every term that
belongs to $a^{\mathrm{t-p}}$ is necessary a product between elements from
the BRST complex of the tensor field with mixed symmetry $\left( 3,1\right) $
and elements from the BRST complex of the abelian form. It was proved in 
\cite{noijhep31} that $a^{\mathrm{t-t}}=0$, while for $a^{\mathrm{p-p}}$ we
will use in a space-time with the dimenssion $D=2p+1$ the solution 
\begin{equation}
a^{\mathrm{p-p}\left( D=2p+1\right) }=\epsilon ^{\mu _{1}...\mu
_{2p+1}}A_{\mu _{1}...\mu _{p}}F_{\mu _{p+1}...\mu _{2p+1}}  \label{tpn2}
\end{equation}%
and in a space-time with the dimenssion $D=3p+2$ the solution%
\begin{equation}
a^{\mathrm{p-p}\left( D=3p+2\right) }=\epsilon ^{\mu _{1}...\mu
_{3p+2}}A_{\mu _{1}...\mu _{p}}F_{\mu _{p+1}...\mu _{2p+1}}F_{\mu
_{2p+2}...\mu _{3p+2}}.  \label{tpn3}
\end{equation}%
(see \cite{citatiicupforme2}).

The component that could generate the cross-interactions, $a^{\mathrm{t-p}}$%
, is determined by the equation%
\begin{equation}
sa^{\mathrm{t-p}}=\partial _{\mu }j^{\mu },  \label{tp18}
\end{equation}%
where $j^{\mu }$ is a local current. We solve the equation (\ref{tp18})
decomposing $a^{\mathrm{t-p}}$ according to the antighost number (we suppose
this decomposition contains a finite number of terms, see \cite{gen1a,gen1b,gen2}),%
\begin{equation}
a^{\mathrm{t-p}}=\sum_{k=0}^{I}a_{k}^{\mathrm{t-p}},\ \mathrm{agh}%
\left( a_{k}^{\mathrm{t-p}}\right) =k,\ \mathrm{gh}\left( a_{k}^{\mathrm{t-p}%
}\right) =0,\ \varepsilon \left( a_{k}^{\mathrm{t-p}}\right) =0.
\label{tp19}
\end{equation}%
If we take account of $s=\delta +\gamma $, the equation (\ref{tp18}) is
equivalent to a chain of equations%
\begin{eqnarray}
\gamma a_{I}^{\mathrm{t-p}} &=&0,  \label{tp20} \\
\delta a_{k}^{\mathrm{t-p}}+\gamma a_{k-1}^{\mathrm{t-p}} &=&\partial _{\mu }%
\overset{\left( k-1\right) }{w}^{\mu },\ I\geq k\geq 1.  \label{tp21}
\end{eqnarray}%
where $\left( \overset{(k)}{w}^{\mu }\right) _{k=\overline{0,I}}$ are local
currents with $\mathrm{agh}\left( \overset{(k)}{w}^{\mu }\right) =k$, while
we have noted by $I$ the greatest antighost number from the decomposition (%
\ref{tp19}) (see \cite{def,gen1a,gen1b} for further details). We say the
chain (\ref{tp20})-(\ref{tp21}) is consistent if all its equations have
solutions (the equations are solved from the greatest antighost number of
the chain, to the zero order). Also, it is possible that two chains of the
type (\ref{tp20})-(\ref{tp21}) (one with the greatest antighost number $I$
and the other with the greatest antighost number $J$, each of this two
chains separately inconsistent and the consistency stoping at same antighost
number for both chains) together to be consistent and to form a solution of
the equation (\ref{tp18}).

The solution of the equation (\ref{tp20}) pertains to $H\left( \gamma
\right) $. The generators of this cohomology associated to the tensor field $%
t_{\lambda \mu \nu |\alpha }$ were calculated in \cite{noijhep31} and these
are all the antifields from the BRST complex of $t_{\lambda \mu \nu |\alpha
} $ (noted with $\Pi ^{\ast \Delta }$), their derivatives, the curvature
tensor $K_{\lambda \mu \nu \xi |\alpha \beta }$ and its derivatives, the
ghosts $\mathcal{F}_{\mu \nu \lambda \rho }\mathcal{\equiv }\partial
_{\lbrack \mu }\eta _{\nu \lambda \rho ]}$ and $C_{\mu }$ ($\mathcal{F}_{\mu
\nu \lambda \rho }\in H^{1}\left( \gamma \right) ,\,C_{\mu }\in H^{3}\left(
\gamma \right) $), while from the BRST complex of the abelian form we have
as generators of $H\left( \gamma \right) $ all the antifields and their
derivatives, the field strength $F_{\mu _{1}...\mu _{p+1}}$ and the ghost $%
\overset{\left( p\right) }{\xi }$ ($\overset{\left( p\right) }{\xi }\in
H^{p}\left( \gamma \right) $). Therefore, up to $\gamma $-exact terms, the
general solution for (\ref{tp20}) has the form%
\begin{equation}
a_{I}^{\mathrm{t-p}}=\alpha _{I}^{\mathrm{t-p}}\left( \left[ \pi ^{\ast
\Theta }\right] ,\left[ K_{\lambda \mu \nu \xi |\alpha \beta }\right] ,\left[
F_{\mu _{1}...\mu _{p+1}}\right] \right) e^{I}\left( \overset{\left(
p\right) }{\xi },C_{\nu },\mathcal{F}_{\lambda \mu \nu \alpha }\right)
\label{tp22}
\end{equation}%
where we have noted by $\pi ^{\ast \Theta }=\left( \Pi ^{\ast \Delta
},A_{\mu _{1}...\mu _{p}}^{\ast },\overset{\left( k\right) }{\xi }_{\mu
_{1}...\mu _{p-k}}^{\ast }\right) $ all the antifields. $\alpha _{I}^{t-p}$
are called \textquotedblleft invariant polynomials\textquotedblright\ (the
invariant polynomials are objects with zero pureghost number and $\gamma $%
-closed) and they introduce into the interactions study the local cohomology
of the Koszul-Tate differential, because for the chain (\ref{tp20})-(\ref%
{tp21}) to have solutions it is necessary the invariant polynomials to be $%
\delta $-closed modulo the exterior space-time differential $d$, 
\begin{equation}
\delta \alpha _{I}^{\mathrm{t-p}}=\partial _{\mu }\beta _{I}^{\mu }.
\label{tp22b}
\end{equation}%
The objects $\beta _{I}^{\mu }$ from the previous formula may be chosen
invariant polynomials (see \cite{gen1a,gen1b,noijhep31}), while about the
trivial solutions $\alpha _{I}^{t-p}=\delta \alpha _{I+1}+\partial _{\mu
}\lambda _{I}^{\mu }$ first it can be proved the objects $\alpha _{I+1}$ and 
$\lambda _{I}^{\mu }$ are invariant polynomials and, second, we can remove
from the first order deformation the terms formed with trivial invariant
polynomials. Hence, we are interested, in fact, about the local cohomology
of the Koszul-Tate differential in the space of the invariant polynomials, $%
H_{I}^{\mathrm{inv}}\left( \delta |d\right) $. If we denote the Cauchy order
of the theory (\ref{tp1}) by \textrm{Ord}, the theorems regarding the local
cohomologies prove that%
\begin{equation}
H_{I}\left( \delta |d\right) =0,\ \mathrm{if\ }I>\mathrm{Ord},
\label{tp23}
\end{equation}%
therefore%
\begin{equation}
H_{I}^{\mathrm{inv}}\left( \delta |d\right) =0,\ \mathrm{if\ }I>\mathrm{%
Ord,}  \label{tp23b}
\end{equation}%
so the first order deformation will contain only terms with the antighost
number less or equal than the Cauchy order. In our theory (\ref{tp1}), if
the degree of the abelian form is $p>3$ the Cauchy order is \textrm{Ord}$%
=p+1 $, else the Cauchy is \textrm{Ord}$=4$. Consequently, for the greatest
antighost number in the first order deformation we will have three main
possibilities, analysed separately in the sequel. In all the cases we will
analyse it is maintained the requirement that the possible interacting
Lagrangian has the maximum derivative order two.

\section{$I\geq 5$}

This case appears if $p\geq 4$, all the cohomological groups $H_{I}^{\mathrm{%
inv}}\left( \delta |d\right) $ being nontrivial for $I\leq p+1$. If $I\geq 5$%
, the last term from the decomposition (\ref{tp19}) will be linear in the
antifield $\overset{\left( I-1\right) }{\xi }^{\ast \mu _{1}...\mu _{p-I+1}}$
from the sector of an abelian form with $p\geq 4$,%
\begin{equation}
a_{I}^{\mathrm{t-p}}=\overset{\left( I-1\right) }{\xi }^{\ast \mu _{1}...\mu
_{p-I+1}}e_{\mu _{1}...\mu _{p-I+1}}^{I}\left( \overset{\left( p\right) }{%
\xi },C_{\nu },\mathcal{F}_{\lambda \mu \nu \alpha }\right) .  \label{tp25}
\end{equation}%
$e_{\mu _{1}...\mu _{p-I+1}}^{I}$ are nontrivial objects from $H\left(
\gamma \right) $, with the pureghost number $I\leq p+1$, ,constructed from
the ghosts $\overset{\left( p\right) }{\xi },$ $C_{\nu },$ $\mathcal{F}%
_{\lambda \mu \nu \alpha }$ and, possibly, the metric and the Levi-Civita
symbols, thus to have $p-I+1$ Lorentz indices. We are interested by the
computation of the cross-interactions, so $a_{I}^{t-p}$ from (\ref{tp25})
has to contain at least one ghost from the sector of the tensor field with
the mixed symmetry $\left( 3,1\right) $. The restriction imposed to the
derivative order constrains $a_{I}^{t-p}$ to depend of one ghost $\mathcal{F}%
_{\lambda \mu \nu \alpha }$, at most. If we want the object $e_{\mu
_{1}...\mu _{p-I+1}}^{I}$ to depend of the ghost $\overset{\left(
p\right) }{\xi }$ from the sector of the abelian form, then the only
possibility is $I=p+1$ and%
\begin{equation}
a_{I}^{\mathrm{t-p}}=f^{\lambda \mu \nu \alpha }\overset{\left( p+1\right) }{%
\xi }^{\ast }\overset{\left( p\right) }{\xi }\mathcal{F}_{\lambda \mu \nu
\alpha },  \label{tpn26}
\end{equation}%
where $f^{\lambda \mu \nu \alpha }$ is constant (because $\mathrm{gh}\left( 
\overset{\left( p\right) }{\xi }C_{\mu }\right) =p+3$, the product $\overset{%
\left( p\right) }{\xi }C_{\mu }$ can not appear in $a_{I}^{t-p}$). It is not
possible to construct the object $a_{I}^{t-p}$ with the form (\ref{tpn26}),
because of the space-time dimenssion, $D\geq 5$.

We remain with two possiblities: either $e_{\mu _{1}...\mu _{p-I+1}}^{I}$ it
is formed only from the ghosts $C_{\mu }$, or from the ghosts $C_{\mu }$ and
one ghost $F_{\lambda \mu \nu \alpha }$. We will analyse each of these
possibilities. In the first case, the general form of $a_{I}^{t-p}$ is%
\begin{equation}
a_{I}^{\mathrm{t-p}}=f^{\mu _{1}...\mu _{p-I+1}||\nu _{1}...\nu _{N}}\overset%
{\left( I-1\right) }{\xi }_{\mu _{1}...\mu _{p-I+1}}^{\ast }C_{\nu
_{1}}...C_{\nu _{N}},  \label{tpn27}
\end{equation}%
where $f^{\mu _{1}...\mu _{p-I+1}||\nu _{1}...\nu _{N}}$ is a constant
tensor, antisymmetric separately in the groups of indices $\mu _{1}...\mu
_{p-I+1}$, respectively $\nu _{1}...\nu _{N}$. Furthermore, the following
conditions are satisfied,%
\begin{equation}
I=3N,5\leq I\leq p+1\Rightarrow N\geq 2.  \label{tpn28}
\end{equation}%
We deduce from (\ref{tpn27}) the equation $\delta a_{I}^{\mathrm{t-p}%
}+\gamma a_{I-1}^{\mathrm{t-p}}=\partial _{\mu }j_{I-1}^{\mu }$ has the
solution%
\begin{eqnarray}
a_{I-1}^{\mathrm{t-p}} &=&-\frac{N\left( p-I+2\right) }{6}f^{\mu _{1}...\mu
_{p-I+1}||\nu _{1}...\nu _{N}}\times   \notag \\
&&\overset{\left( I-2\right) }{\xi }_{\ \ \ \ \ \mu _{1}...\mu
_{p-I+1}}^{\ast \lambda }C_{\nu _{1}}...C_{\nu _{N-1}}\mathcal{C}_{\lambda
\nu _{N}}^{^{\prime }},  \label{tpn29}
\end{eqnarray}%
while the equation $\delta a_{I-1}^{\mathrm{t-p}}+\gamma a_{I-2}^{\mathrm{t-p%
}}=\partial _{\mu }j_{I-2}^{\mu }$ has the solution%
\begin{eqnarray}
a_{I-2}^{\mathrm{t-p}} &=&\frac{N\left( p-I+2\right) \left( p-I+3\right) }{12%
}f^{\mu _{1}...\mu _{p-I+1}||\nu _{1}...\nu _{N}}\overset{\left( I-3\right) }%
{\xi }_{\ \ \ \ \ \ \mu _{1}...\mu _{p-I+1}}^{\ast \nu \lambda }\times  
\notag \\
&&\left( \frac{N-1}{6}C_{\nu _{1}}...C_{\nu _{N-2}}\mathcal{C}_{\nu \nu
_{N-1}}^{^{\prime }}\mathcal{C}_{\lambda \nu _{N}}^{^{\prime }}-C_{\nu
_{1}}...C_{\nu _{N-1}}\mathcal{G}_{\nu \lambda |\nu _{N}}^{^{\prime
}}\right) .  \label{tpn30}
\end{eqnarray}%
The action of $\delta $ on $a_{I-2}^{\mathrm{t-p}}$ from (\ref{tpn30}) is%
\begin{eqnarray}
\delta a_{I-2}^{\mathrm{t-p}} &=&\partial _{\mu }\rho _{I-3}^{\mu }+\gamma
\left\{ \frac{N\left( p-I+2\right) \left( p-I+3\right) \left( p-I+4\right) }{%
12}f^{\mu _{1}...\mu _{p-I+1}||\nu _{1}...\nu _{N}}\times \right.   \notag \\
&&\overset{\left( I-4\right) }{\xi }_{\ \ \ \ \ \ \ \ \mu _{1}...\mu
_{p-I+1}}^{\ast \mu \nu \lambda }\left[ \frac{N-1}{6}\left( \frac{N-2}{18}%
C_{\nu _{1}}...C_{\nu _{N-3}}\mathcal{C}_{\mu \nu _{N-2}}^{^{\prime }}%
\mathcal{C}_{\nu \nu _{N-1}}^{^{\prime }}\mathcal{C}_{\lambda \nu
_{N}}^{^{\prime }}+\right. \right.   \notag \\
&&\left. \left. C_{\nu _{1}}...C_{\nu _{N-2}}\mathcal{C}_{\mu \nu
_{N-1}}^{^{\prime }}\mathcal{G}_{\nu \lambda |\nu _{N}}^{^{\prime }}\right)
+\left. \frac{1}{3}C_{\nu _{1}}...C_{\nu _{N-1}}t_{\mu \nu \lambda |\nu _{N}}%
\right] \right\} +  \notag \\
&&\frac{N\left( p-I+2\right) \left( p-I+3\right) \left( p-I+4\right) }{12}%
f^{\mu _{1}...\mu _{p-I+1}||\nu _{1}...\nu _{N}}\times   \notag \\
&&\overset{\left( I-4\right) }{\xi }_{\ \ \ \ \ \ \ \ \mu _{1}...\mu
_{p-I+1}}^{\ast \mu \nu \lambda }C_{\nu _{1}}...C_{\nu _{N-1}}\mathcal{F}%
_{\mu \nu \lambda \nu _{N}}.  \label{tpn31}
\end{eqnarray}%
(Remark about the convention we use for the indices: if $N=2$ the term
containing the product $C_{\nu _{1}}...C_{\nu _{N-3}}$ does not appear
anymore, while the term with the product $C_{\nu _{1}}...C_{\nu _{N-2}}$
will not contain anymore this product.) The last term from the expression (%
\ref{tpn31}) of $\delta a_{I-2}^{\mathrm{t-p}}$ pertains to $H\left( \gamma
\right) $ and the equation $\delta a_{I-2}^{\mathrm{t-p}}+\gamma a_{I-3}^{%
\mathrm{t-p}}=\partial _{\mu }j_{I-3}^{\mu }$ has no solution.

The second possibility we analyse for $5\leq I\leq p+1$ is that $a_{I}^{%
\mathrm{t-p}}$ has the form%
\begin{equation}
a_{I}^{\mathrm{t-p}}=f^{\mu _{1}...\mu _{p-I+1}||\lambda \mu \nu \rho ||\nu
_{1}...\nu _{M}}\overset{\left( I-1\right) }{\xi }_{\mu _{1}...\mu
_{p-I+1}}^{\ast }\mathcal{F}_{\lambda \mu \nu \rho }C_{\nu _{1}}...C_{\nu
_{M}},  \label{tpn32}
\end{equation}%
where $f^{\mu _{1}...\mu _{p-I+1}||\lambda \mu \nu \rho ||\nu _{1}...\nu
_{M}}$ is a constant tensor, antisymmetric separately in the three groups of
indices ($\mu _{1}...\mu _{p-I+1}$, $\lambda \mu \nu \rho $, respectively $%
\nu _{1}...\nu _{M}$) 
\begin{equation}
I=3M+1,5\leq I\leq p+1\Rightarrow M\geq 2.  \label{tpn33}
\end{equation}%
From (\ref{tpn32}), it follows the equation $\delta a_{I}^{%
\mathrm{t-p}}+\gamma a_{I-1}^{\mathrm{t-p}}=\partial _{\mu }j_{I-1}^{\mu }$
has the solution 
\begin{eqnarray}
a_{I-1}^{\mathrm{t-p}} &=&\left( p-I+2\right) f^{\mu _{1}...\mu
_{p-I+1}||\lambda \mu \nu \rho ||\nu _{1}...\nu _{M}}\overset{\left(
I-2\right) }{\xi }_{\ \ \ \ \mu _{1}...\mu _{p-I+1}}^{\ast \alpha }\times  
\notag \\
&&\left( \frac{\left( -1\right) ^{M}}{3}\partial _{\left[ \lambda \right.
}t_{\left. \mu \nu \rho \right] |\alpha }C_{\nu _{1}}...C_{\nu _{M}}-\frac{M%
}{6}\mathcal{F}_{\lambda \mu \nu \rho }C_{\nu _{1}}...C_{\nu _{M-1}}\mathcal{%
C}_{\alpha \nu _{M}}^{^{\prime }}\right) .  \label{tpn34}
\end{eqnarray}%
The action of $\delta $ on $a_{I-1}^{\mathrm{t-p}}$ is 
\begin{eqnarray}
\delta a_{I-1}^{\mathrm{t-p}} &=&\partial _{\mu }\rho _{I-2}^{\mu }+\gamma
\left\{ \left( p-I+2\right) \left( p-I+3\right) \times \right.   \notag \\
&&\left. f^{\mu _{1}...\mu _{p-I+1}||\lambda \mu \nu \rho ||\nu _{1}...\nu
_{M}}\overset{\left( I-3\right) }{\xi }_{\ \ \ \ \ \ \mu _{1}...\mu
_{p-I+1}}^{\ast \alpha \beta }\times \right.   \notag \\
&&\left[ \frac{\left( -1\right) ^{M}}{18}\partial _{\left[ \lambda \right.
}t_{\left. \mu \nu \rho \right] |\alpha }C_{\nu _{1}}...C_{\nu _{M-1}}%
\mathcal{C}_{\beta \nu _{M}}^{^{\prime }}+\frac{M\left( M-1\right) }{6}%
\mathcal{F}_{\lambda \mu \nu \rho }C_{\nu _{1}}...C_{\nu _{M-2}}\mathcal{C}%
_{\beta \nu _{M-1}}^{^{\prime }}\mathcal{C}_{\alpha \nu _{M}}^{^{\prime
}}+\right.   \notag \\
&&\left. \left. \frac{M}{12}\mathcal{F}_{\lambda \mu \nu \rho }C_{\nu
_{1}}...C_{\nu _{M-1}}\mathcal{G}_{\beta \alpha |\nu _{M}}^{^{\prime }}%
\right] \right\} +  \notag \\
&&\frac{\left( -1\right) ^{M+1}}{6}\left( p-I+2\right) \left( p-I+3\right)
f^{\mu _{1}...\mu _{p-I+1}||\lambda \mu \nu \rho ||\nu _{1}...\nu
_{M}}\times   \notag \\
&&\overset{\left( I-3\right) }{\xi }_{\ \ \ \ \ \ \mu _{1}...\mu
_{p-I+1}}^{\ast \alpha \beta }K_{\lambda \mu \nu \rho |\beta \alpha }C_{\nu
_{1}}...C_{\nu _{M}},  \label{tpn35}
\end{eqnarray}%
and the equation $\delta a_{I-1}^{\mathrm{t-p}}+\gamma a_{I-2}^{\mathrm{t-p}%
}=\partial _{\mu }j_{I-2}^{\mu }$ has no solution.

\section{$0<I\leq 4$}

For $I\leq 4$, the cohomological group $H_{I}\left( \delta |d\right) $ may
have, depending on the degree of the abelian form, two types of generators:
the antifields with antighost number $I$ from the sector of the tensor field
with the mixed symmetry $\left( 3,1\right) $ $t_{\lambda \mu \nu |\alpha }$
and the antifields with antighost number $I$, if these exist, from the
sector of the abelian form. The equation (\ref{tp20}) has two independent
solutions, one linear in the antifields from the sector of the tensor field $%
t_{\lambda \mu \nu |\alpha }$ and the other linear in the antifields from
the sector of the abelian form. We will analyse separately the chains of
equations of the type (\ref{tp20})-(\ref{tp21}), starting from each of these
two independent solutions. The consistency for each of these chains goes
separately until the antighost number one, and if both chains are \textbf{%
incosistent in order one, }it is possible that \textbf{together to be
consistent also in the zero order}.

\subsection{Invariant polynomials generated by the antifields from the
sector $\left( 3,1\right) $}

The necessary condition to generate the cross-interactions is that $a_{I}^{%
\mathrm{t-p}}$ to mix the objects from the sector $\left( 3,1\right) $ and
the objects from the sector of the abelian form, so we need at least the
form $\left( a_{I}^{\mathrm{t-p}}\right) _{I\leq 4}=\left( \mathrm{%
antifields\ }\left( 3,1\right) \right) \times \left( \mathrm{form\ ghosts}%
\right) $. The only ghost corresponding to the abelian form nontrivial in $%
H\left( \gamma \right) $ is$\ \overset{\left( p\right) }{\xi }$ and it must
to have the pureghost number less or equal than four. It follows that in
this case we can take into account only the abelian forms with the degrees $%
p\leq 4$. In the sequel, the analysis will depend on the degree $p$ of the
abelian form.

If $p=4$ and $I=4$ the general form of the last component from (\ref{tp19})
is 
\begin{equation}
a_{4}^{\mathrm{t-p}}=f^{\mu }C_{\mu }^{\ast }\overset{\left( 4\right) }{\xi }
\label{tp28}
\end{equation}%
and, because $f^{\mu }$ is constant, it can not be constructed.

\subsubsection{3-forms, $p=3$}

If $p=3$, $I=4$ the last component from (\ref{tp19}) can be constructed in a
space-time with the dimension $D=5,$ 
\begin{equation}
a_{4}^{\mathrm{t-p}\left( D=5\right) }=\epsilon ^{\mu _{1}...\mu _{5}}C_{\mu
_{1}}^{\ast }\partial _{\lbrack \mu _{2}}\eta _{\mu _{2}\mu _{3}\mu _{4}]}%
\overset{\left( 3\right) }{\xi }.  \label{tp29}
\end{equation}%
The equation $\delta a_{4}^{t-p\left( D=5\right) }+\gamma a_{3}^{t-p\left(
D=5\right) }=\partial ^{\mu }\overset{\left( 3\right) }{j}_{\mu }$ has the
solution%
\begin{equation}
a_{3}^{\mathrm{t-p}\left( D=5\right) }=-2\epsilon ^{\mu _{1}...\mu _{5}}%
\mathcal{C}_{\,\,\,\,\,\,\,\,\,\,\,\,\mu _{1}}^{^{\prime }\ast \lambda
}\left( \partial _{\lbrack \mu _{2}}t_{\mu _{3}\mu _{4}\mu _{5}]|\lambda }%
\overset{\left( 3\right) }{\xi }-3\partial _{\lbrack \mu _{2}}\eta _{\mu
_{3}\mu _{4}\mu _{5}]}\overset{\left( 2\right) }{\xi }_{\lambda }\right) ,
\label{tp30}
\end{equation}%
but the consistency stops here (the equation $\delta a_{3}^{\mathrm{t-p}%
\left( D=5\right) }+\gamma a_{2}^{\mathrm{t-p}\left( D=5\right) }=\partial
^{\mu }\overset{\left( 2\right) }{j}_{\mu }$ has no solution $a_{2}^{\mathrm{%
t-p}\left( D=5\right) }$), because in 
\begin{eqnarray}
\delta a_{3}^{\mathrm{t-p}\left( D=5\right) } &=&\partial ^{\lambda }\rho
_{\lambda }+\gamma \left( \epsilon ^{\mu _{1}...\mu _{5}}\mathcal{G}_{\rho
\lambda |\mu _{1}}^{^{\prime }\ast }\left( -4\partial _{\lbrack \mu
_{2}}t_{\mu _{3}\mu _{4}\mu _{5}]}|^{\lambda }\overset{\left( 2\right) }{\xi 
}^{\rho }+6\partial _{\lbrack \mu _{2}}\eta _{\mu _{3}\mu _{4}\mu _{5}]}%
\overset{\left( 1\right) }{\xi }^{\rho \lambda }\right) \right) -  \notag \\
&&2\epsilon ^{\mu _{1}...\mu _{5}}\mathcal{G}_{\rho \lambda |\mu
_{1}}^{^{\prime }\ast }\partial _{\lbrack \mu _{2}}t_{\mu _{3}\mu _{4}\mu
_{5}]}|^{[\lambda ,\rho ]}\overset{\left( 3\right) }{\xi },  \label{tp31}
\end{eqnarray}%
the last term is nontrivial in $H^{3}\left( \gamma \right) $, so $\delta
a_{3}^{\mathrm{t-p}\left( D=5\right) }\neq \partial ^{\mu }\overset{\left(
2\right) }{j}_{\mu }+\left( \gamma \mathrm{-exact}\right) $.

If $p=3$, $I=3$ we can construct the last component from (\ref{tp19}) in a
space-time with the dimension $D\geq 5$,%
\begin{equation}
a_{3}^{\mathrm{t-p}}=\mathcal{C}_{\,\,\,\,\,\,\,\,\,\mu }^{\ast \mu }\overset%
{\left( 3\right) }{\xi }.  \label{tp32}
\end{equation}%
The equation $\delta a_{3}^{\mathrm{t-p}}+\gamma a_{2}^{\mathrm{t-p}%
}=\partial ^{\mu }\overset{\left( 2\right) }{j}_{\mu }$ has the solution%
\begin{equation}
a_{2}^{\mathrm{t-p}}=-2\mathcal{G}^{\ast \lambda \mu }|_{\mu }\overset{%
\left( 2\right) }{\xi }_{\lambda },  \label{tp33}
\end{equation}%
while for $\delta a_{2}^{\mathrm{t-p}}+\gamma a_{1}^{\mathrm{t-p}}=\partial
^{\mu }\overset{\left( 1\right) }{j}_{\mu }$ we find the solution%
\begin{equation}
a_{1}^{\mathrm{t-p}}=3t^{\ast \rho \lambda \mu }|_{\mu }\overset{\left(
1\right) }{\xi }_{\rho \lambda }.  \label{tp34}
\end{equation}%
From the last formula it follows 
\begin{equation}
\delta a_{1}^{\mathrm{t-p}}=3T^{\rho \lambda \mu }|_{\mu }\overset{\left(
1\right) }{\xi }_{\rho \lambda },  \label{tp35}
\end{equation}%
where $T^{\rho \lambda \mu |\alpha }$ are the functions appearing in the
field equations of $t_{\lambda \mu \nu |\rho }$ (see (\ref{t26})). These
functions may be expressed using $F^{\lambda \mu \nu \xi |\alpha \beta }$
(see (\ref{t26})), following the contraction $T^{\rho \lambda \mu }|_{\mu
}=\left( \frac{D}{2}-2\right) F^{\rho \lambda \alpha \beta }|_{\alpha \beta
} $, which we introduce in (\ref{tp35}) and we obtain%
\begin{equation}
\delta a_{1}^{\mathrm{t-p}}=\partial ^{\lambda }\rho _{\lambda }-3\left( 
\frac{D}{2}-2\right) \partial ^{\lbrack \rho }t^{\lambda \alpha \beta
]}|_{\beta }\cdot \gamma A_{\rho \lambda \alpha }.  \label{tp36}
\end{equation}%
The last term from (\ref{tp36}) is not $\gamma $-closed, so $a_{1}^{t-p}$
from (\ref{tp34}) is not consistent (however, we will see later the chain of
the type (\ref{tp20})-(\ref{tp21}) starting from the component (\ref{tp32})
will be consistent \textbf{together }with a chain depending on the
antifields from the sector of the 3-form).

For $p=3$ and $I\leq 2$, the last component from (\ref{tp19}) can not be
constructed.

\subsubsection{2-forms, $p=2$}

If $p=2,$ $I=4$ the general form of the last component from (\ref{tp19}) is%
\begin{equation}
a_{4}^{t-p}=f^{\mu _{1}...\mu _{9}}C_{\mu _{1}}^{\ast }\partial _{\lbrack
\mu _{2}}\eta _{\mu _{3}\mu _{4}\mu _{5}]}\partial _{\lbrack \mu _{6}}\eta
_{\mu _{7}\mu _{8}\mu _{9}]}\overset{\left( 2\right) }{\xi },  \label{tp37}
\end{equation}%
and can not be constructed, because of the restriction on the derivative
order.

If $p=2$, $I=3$ we can construct the last component from (\ref{tp19}) in a
space-time with the dimension $D=6$, 
\begin{equation}
a_{3}^{t-p\left( D=6\right) }=\epsilon ^{\mu _{1}...\mu _{6}}C_{\mu _{1}\mu
_{2}}^{\ast }\partial _{\lbrack \mu _{3}}\eta _{\mu _{4}\mu _{5}\mu _{6}]}%
\overset{\left( 2\right) }{\xi }.  \label{tp38}
\end{equation}%
The equation $\delta a_{3}^{t-p\left( D=6\right) }+\gamma a_{2}^{t-p\left(
D=6\right) }=\partial ^{\mu }\overset{\left( 2\right) }{j}_{\mu }$ has the
solution%
\begin{equation}
a_{2}^{\mathrm{t-p}\left( D=6\right) }=3\epsilon ^{\mu _{1}...\mu
_{6}}\left( \mathcal{G}_{\mu _{1}\mu _{2}}^{\ast }|^{\lambda }-\frac{1}{2}%
\eta _{\mu _{1}\mu _{2}}^{\ast \,\,\,\,\,\,\,\,\,\,\,\,\,\,\lambda }\right)
\left( -\frac{1}{3}\partial _{\lbrack \mu _{3}}t_{\mu _{4}\mu _{5}\mu
_{6}]|\lambda }\overset{\left( 2\right) }{\xi }+\partial _{\lbrack \mu
_{3}}\eta _{\mu _{4}\mu _{5}\mu _{6}]}\overset{\left( 1\right) }{\xi }%
_{\lambda }\right) ,  \label{tp39}
\end{equation}%
but the consistency ends here, because in%
\begin{eqnarray}
\delta a_{2}^{\mathrm{t-p}\left( D=6\right) } &=&\partial ^{\lambda }\rho
_{\lambda }+\gamma \left( 9\epsilon ^{\mu _{1}...\mu _{6}}t_{\mu _{1}\mu _{2}%
\left[ \rho |\lambda \right] }^{\ast }\left( -\frac{1}{3}\partial _{\lbrack
\mu _{3}}t_{\mu _{4}\mu _{5}\mu _{6}]}|^{\lambda }\overset{\left( 1\right) }{%
\xi }^{\rho }+\frac{1}{2}\partial _{\lbrack \mu _{3}}\eta _{\mu _{4}\mu
_{5}\mu _{6}]}A^{\rho \lambda }\right) \right) +  \notag \\
&&\frac{1}{6}\epsilon ^{\mu _{1}...\mu _{6}}t_{\mu _{1}\mu _{2}\left[ \rho
|\lambda \right] }^{\ast }\partial _{\lbrack \mu _{3}}t_{\mu _{4}\mu _{5}\mu
_{6}]}|^{\left[ \rho ,\lambda \right] }\overset{\left( 2\right) }{\xi }
\label{tp40}
\end{eqnarray}%
the last term is nontrivial in $H^{3}\left( \gamma \right) $ and the
equation $\delta a_{2}^{t-p\left( D=6\right) }+\gamma a_{1}^{t-p\left(
D=6\right) }=\partial ^{\mu }\overset{\left( 1\right) }{j}_{\mu }$ has no $%
a_{1}^{t-p\left( D=6\right) }$ solution.

If $p=2$, $I=2$, the last component from (\ref{tp19}) is (in a space time
with the dimension $D\geq 5$) 
\begin{equation}
a_{2}^{\mathrm{t-p}}=\eta ^{\ast \mu \nu \lambda }\partial _{\lbrack \mu
}A_{\nu \lambda ]}\overset{\left( 2\right) }{\xi }.  \label{tp41}
\end{equation}%
The previous object is not consistent, because in 
\begin{equation}
\delta a_{2}^{\mathrm{t-p}}=\partial _{\lambda }\rho ^{\lambda }+\gamma
\left( t^{\ast \mu \nu \lambda |\rho }\partial _{\lbrack \mu }A_{\nu \lambda
]}\overset{\left( 1\right) }{\xi }_{\rho }\right) +t^{\ast \mu \nu \lambda
|\rho }\partial _{\rho }\partial _{\lbrack \mu }A_{\nu \lambda ]}\overset{%
\left( 2\right) }{\xi },  \label{tp42}
\end{equation}%
the last term is not $\gamma $-exact.

\subsubsection{Vector fields, $p=1$}

This case was analysed in \cite{31vect}.

\subsection{Invariant polynomials generated by the antifields from the
sector of an abelian form}

The condition that the solution of the equation $\gamma a_{I}=0$ to mix the
objects from the BRST complex $\left( 3,1\right) $ and the objects from the
BRST complex of an abelian form forces the invariant polynomials linear in
the antifields from the sector of an abelian form, $a_{I}$ to depend on the
ghosts from the sector $\left( 3,1\right) $. It is simple to prove $a_{I}$
do not depend in this case on the ghosts of the abelian form, namely we
don't have solutions of the type $a_{I}=\mathrm{form\ antifields}\times 
\mathrm{form\ ghosts}\times \mathrm{ghosts\,}\left( 3,1\right) $. Such a
solution would have the general form 
\begin{equation}
a_{I}=f^{\mu _{1}\mu _{2}\mu _{3}\mu _{4}}\overset{\left( p\right) \ast }{%
\xi }\overset{\left( p\right) }{\xi }\partial _{\lbrack \mu _{1}}\eta _{\mu
_{2}\mu _{3}\mu _{4}]},\;I=p+1,  \label{tp61}
\end{equation}%
that can not constructed concretely, because of the condition on the
space-time dimension $D\geq 5$. Therefore, we will search for the component
with the greatest antighost number from (\ref{tp19}) the solution of the
type $a_{I}=\mathrm{form\ antifields}\times \mathrm{ghosts\,}\left(
3,1\right) $. Our analysis will not depend so much on the degree of the
abelian form, as in the case when the invariant polynomials where generated
by the antifields from the sector $\left( 3,1\right) $.

\subsubsection{$I=4$}

If $I=4$, $p\geq 3$ we can construct the last component from (\ref{tp19}) in
a space time with the dimension $D=p+2\geq 5$, 
\begin{equation}
a_{4}^{\mathrm{t-p}}=\epsilon ^{\mu _{1}...\mu _{p+2}}\overset{\left(
3\right) }{\xi }_{\mu _{1}...\mu _{p-3}}^{\ast }C_{\mu _{p-2}}\partial
_{\lbrack \mu _{p-1}}\eta _{\mu _{p}\mu _{p+1}\mu _{p+2}]}.  \label{tp62}
\end{equation}%
The equation $\delta a_{4}^{t-p}+\gamma a_{3}^{t-p}=\partial _{\mu }\overset{%
\left( 2\right) }{j}^{\mu }$ has the solution%
\begin{eqnarray}
a_{3}^{\mathrm{t-p}} &=&-\left( p-2\right) \epsilon ^{\mu _{1}...\mu _{p+2}}%
\overset{\left( 2\right) }{\xi }_{\lambda \mu _{1}...\mu _{p-3}}^{\ast
}\left( \frac{1}{6}\mathcal{C}_{\,\,\,\,\,\,\,\,\mu _{p-2}}^{^{\prime
}\lambda }\partial _{\lbrack \mu _{p-1}}\eta _{\mu _{p}\mu _{p+1}\mu
_{p+2}]}+\right.   \notag \\
&&\left. \frac{1}{3}C_{\mu _{p-2}}\partial _{\lbrack \mu _{p-1}}t_{\mu
_{p}\mu _{p+1}\mu _{p+2}]}|^{\lambda }\right) ,  \label{tp63}
\end{eqnarray}%
from which it follows%
\begin{eqnarray}
\delta a_{3}^{\mathrm{t-p}} &=&\partial ^{\lambda }j_{\lambda }-\gamma \left[
\left( p-2\right) \left( p-1\right) \epsilon ^{\mu _{1}...\mu _{p+2}}\overset%
{\left( 1\right) }{\xi }_{\rho \lambda \mu _{1}...\mu _{p-3}}^{\ast }\times
\right.   \notag \\
&&\left. \left( \frac{1}{12}\mathcal{G}^{^{\prime }\rho \lambda }|_{\mu
_{p-2}}\partial _{\lbrack \mu _{p-1}}\eta _{\mu _{p}\mu _{p+1}\mu _{p+2}]}+%
\frac{1}{18}\mathcal{C}_{\,\,\,\,\,\,\,\,\mu _{p-2}}^{^{\prime }\lambda
}\partial _{\lbrack \mu _{p-1}}t_{\mu _{p}\mu _{p+1}\mu _{p+2}]}|^{\rho
}\right) \right] -  \notag \\
&&\frac{\left( p-2\right) \left( p-1\right) }{6}\epsilon ^{\mu _{1}...\mu
_{p+2}}\overset{\left( 1\right) }{\xi }_{\rho \lambda \mu _{1}...\mu
_{p-3}}^{\ast }C_{\mu _{p-2}}\partial _{\lbrack \mu _{p-1}}t_{\mu _{p}\mu
_{p+1}\mu _{p+2}]}|^{\left[ \lambda ,\rho \right] }.  \label{tp64}
\end{eqnarray}%
The last term from the previous expression is nontrivial in $H^{3}\left(
\gamma \right) $, so $\delta a_{3}\neq \partial ^{\lambda }\rho _{\lambda
}+\left( \gamma -exact\right) $ and the consistency of the object $%
a_{4}^{t-p}$ from (\ref{tp62}) ends here.

\subsubsection{$I=3$}

If $I=3$, the general form of the last component from (\ref{tp19}) is%
\begin{equation}
a_{3}^{\mathrm{t-p}}=f^{\mu _{1}...\mu _{p-1}}\overset{\left( 2\right) }{\xi 
}_{\mu _{1}...\mu _{p-2}}^{\ast }C_{\mu _{p-1}},  \label{tp65}
\end{equation}%
and it can be constructed concretely only for $p=3$,%
\begin{equation}
a_{3}^{\mathrm{t-p}}=\overset{\left( 2\right) }{\xi }^{\ast \mu }C_{\mu }.
\label{tpn66}
\end{equation}%
It follows from (\ref{tpn66}) the solution of the equation $\delta a_{3}^{%
\mathrm{t-p}}+\gamma a_{2}^{\mathrm{t-p}}=\partial _{\mu }j_{2}^{\mu }$,%
\begin{equation}
a_{2}^{\mathrm{t-p}}=\overset{\left( 1\right) }{\xi }^{\ast \nu \mu }C_{\nu
\mu }.  \label{tpn67}
\end{equation}%
Next, the equation $\delta a_{2}^{\mathrm{t-p}}+\gamma a_{1}^{\mathrm{t-p}%
}=\partial _{\mu }j_{1}^{\mu }$ has the solution%
\begin{equation}
a_{1}^{\mathrm{t-p}}=-\frac{1}{2}A^{\ast \lambda \nu \mu }\eta _{\lambda \nu
\mu },  \label{tpn68}
\end{equation}%
and the action of $\delta $ on $a_{1}^{\mathrm{t-p}}$ from (\ref{tpn68}) is%
\begin{equation}
\delta a_{1}^{\mathrm{t-p}}=\partial _{\mu }\rho ^{\mu }+\gamma \left( \frac{%
1}{6\cdot 3!}A^{\lambda \nu \mu }\partial _{\left[ \rho \right. }t_{\left.
\lambda \nu \mu \right] }|^{\rho }\right) -\frac{1}{6\cdot 3!}\gamma
A^{\lambda \nu \mu }\partial _{\left[ \rho \right. }t_{\left. \lambda \nu
\mu \right] }|^{\rho }  \label{tpn69}
\end{equation}%
and the consistency stops here, because the last term from (\ref{tpn69}) is
not $\gamma $-exact (Remark: still, we will use this chain, starting from (%
\ref{tpn66}), together with the chain starting from (\ref{tp32}) to obtain a
solution for the equation (\ref{tp18})).

\subsubsection{$I=2$}

If $I=3$, the general form of the last component from (\ref{tp19}) is 
\begin{equation}
a_{2}^{t-p}=f^{\mu _{1}...\mu _{p+7}}\overset{\left( 1\right) }{\xi }_{\mu
_{1}...\mu _{p-1}}^{\ast }\partial _{\lbrack \mu _{p}}\eta _{\mu _{p+1}\mu
_{p+2}\mu _{p+3}]}\partial _{\lbrack \mu _{p+4}}\eta _{\mu _{p+5}\mu
_{p+6}\mu _{p+7}]},  \label{tp66}
\end{equation}%
and can not be constructed because of the condition on the derivative order
(if it was consistent, the component $a_{2}^{t-p}$ from the previous formula
would produce in a possible $a_{0}^{t-p}$ terms with three derivatives).

\subsubsection{$I=1$}

If $I=1$, the decomposition (\ref{tp19}) has two terms,%
\begin{equation}
a^{\mathrm{t-p}}=a_{0}^{\mathrm{t-p}}+a_{1}^{\mathrm{t-p}}.  \label{tp67}
\end{equation}%
The last term from (\ref{tp67})\ has the properties $\mathrm{agh}\left(
a_{1}^{t-p}\right) =1$, $\mathrm{pgh}\left( a_{1}^{t-p}\right) =1$ and
satifies the equation $\gamma a_{1}^{t-p}=0$, with the solution%
\begin{equation}
a_{1}^{\mathrm{t-p}}=\epsilon ^{\mu _{1}...\mu _{p+4}}A_{\mu _{1}...\mu
_{p}}^{\ast }\partial _{\lbrack \mu _{p+1}}\eta _{\mu _{p+2}\mu _{p+3}\mu
_{p+4}]},  \label{tp68}
\end{equation}%
for every value of the abelian form degree ($p\geq 1$), in a space time with
the dimension $D=p+4\geq 5$. The equation $\delta a_{1}^{\mathrm{t-p}%
}+\gamma a_{0}^{\mathrm{t-p}}=\partial _{\mu }\overset{\left( 0\right) }{j}%
^{\mu }$ has the solution 
\begin{equation}
a_{0}^{\mathrm{t-p}}=\frac{1}{3p!}\epsilon ^{\mu _{1}...\mu
_{p+4}}F_{\lambda \mu _{1}...\mu _{p}}\partial _{\lbrack \mu _{p+1}}t_{\mu
_{p+2}\mu _{p+3}\mu _{p+4}]}|^{\lambda }.  \label{tp69}
\end{equation}%
In this moment we have discovered, starting from (\ref{tp67}), the first
order deformation of the solution of the master equation for the theory (\ref%
{tp1}), 
\begin{eqnarray}
S_{1} &=&\int d^{D}x\ \epsilon ^{\mu _{1}...\mu _{p+4}}\left( A_{\mu
_{1}...\mu _{p}}^{\ast }\partial _{\lbrack \mu _{p+1}}\eta _{\mu _{p+2}\mu
_{p+3}\mu _{p+4}]}+\right.   \notag \\
&&\left. \frac{1}{3p!}F_{\lambda \mu _{1}...\mu _{p}}\partial _{\lbrack \mu
_{p+1}}t_{\mu _{p+2}\mu _{p+3}\mu _{p+4}]}|^{\lambda }\right) .  \label{tp70}
\end{eqnarray}%
[Remark: if $I=1$, $p\geq 3$ we could construct, apparently, in a space-time
with the dimension $D=p+2\geq 5$ 
\begin{equation}
a_{1}^{\mathrm{t-p}}=\epsilon ^{\mu _{1}...\mu _{p+2}}A_{\mu _{1}...\mu
_{p-1}\lambda }^{\ast }\partial _{\lbrack \mu _{p}}\eta _{\mu _{p+1}\mu
_{p+2}\rho ]}\sigma ^{\lambda \rho },  \label{tp76}
\end{equation}%
but 
\begin{equation}
\epsilon ^{\mu _{1}...\mu _{p+2}}A_{\mu _{1}...\mu _{p-1}\lambda }^{\ast
}\partial _{\lbrack \mu _{p}}\eta _{\mu _{p+1}\mu _{p+2}\rho ]}\sigma
^{\lambda \rho }=-\frac{3}{p}\epsilon ^{\mu _{1}...\mu _{p+2}}A_{\mu
_{1}...\mu _{p}}^{\ast }\partial _{\lbrack \mu _{p+1}}\eta _{\mu
_{p+2}\lambda \rho ]}\sigma ^{\lambda \rho }=0.  \label{tp77}
\end{equation}%
]

\section{$I=0$}

In this case the first order deformation contains only a component with
antighost number zero, $a^{\mathrm{t-p}}=a_{0}^{\mathrm{t-p}}$, and we have
to solve the equation%
\begin{equation}
\gamma a_{0}^{\mathrm{t-p}}=\partial _{\mu }m_{\mathrm{t-p}}^{\mu },
\label{tpno1}
\end{equation}%
where $a_{0}^{\mathrm{t-p}}$ depends only on the tensor field $\left(
3,1\right) $ and the abelian form (because \textrm{pgh}$a_{0}^{\mathrm{t-p}%
}=0$), and $m_{\mathrm{t-p}}^{\mu }\neq 0$ (the case $m_{\mathrm{t-p}}^{\mu
}=0$ is easily eliminated, see \cite{noijhep31,31vect}). To find the
solutions for (\ref{tpno1}), we shall adopt the method used in \cite%
{noijhep31} slightly modified for cross-interactions. We introduce two
counting operators, one for the mixed symmetry tensor field and its
derivatives,%
\begin{equation}
N^{\left( t\right) }=t_{\lambda \mu \nu |\rho }\frac{\partial }{\partial
t_{\lambda \mu \nu |\rho }}+\sum\limits_{k>0}\partial _{\mu _{1}}...\partial
_{\mu _{k}}t_{\lambda \mu \nu |\rho }\frac{\partial }{\partial \left(
\partial _{\mu _{1}}...\partial _{\mu _{k}}t_{\lambda \mu \nu |\rho }\right) 
},  \label{tpno2}
\end{equation}%
and the other for the abelian form and its derivatives,%
\begin{equation}
N^{\left( A\right) }=A_{\mu _{1}...\mu _{p}}\frac{\partial }{\partial A_{\mu
_{1}...\mu _{p}}}+\sum\limits_{k>0}\partial _{\nu _{1}}...\partial _{\nu
_{k}}A_{\mu _{1}...\mu _{p}}\frac{\partial }{\partial \left( \partial _{\nu
_{1}}...\partial _{\nu _{k}}A_{\mu _{1}...\mu _{p}}\right) }.  \label{tpno3}
\end{equation}%
The solution for (\ref{tpno1}) is written as a sum of eigen solutions for
the counting operators (\ref{tpno2})-(\ref{tpno3}), $a_{0}^{\mathrm{t-p}%
}=\sum\limits_{k,l}a_{kl}$ ($k,l\in \mathbb{N}$, $N^{\left( t\right)
}a_{kl}=ka_{kl}$,$\ N^{\left( A\right) }a_{kl}=la_{kl}$), and it can be
proved that every component $a_{kl}$ from $a_{0}^{\mathrm{t-p}}$ is
separately a solution for (\ref{tpno1}). Therefore, we search a solution $%
a_{0}^{\mathrm{t-p}}\equiv a_{kl}$ for the equation (\ref{tpno1}) that is in
the same time an eigen solution for $N^{\left( t\right) }$ and for $%
N^{\left( A\right) }$. We denote the functional derivatives of $a_{0}^{%
\mathrm{t-p}}$ by%
\begin{equation}
D^{\mu \nu \lambda |\rho }=\frac{\delta a_{0}^{\mathrm{t-A}}}{\delta t_{\mu
\nu \lambda |\rho }},\;D^{\mu _{1}...\mu _{p}}=\frac{\delta a_{0}^{\mathrm{%
t-A}}}{\delta A_{\mu _{1}...\mu _{p}}}  \label{tpno4}
\end{equation}%
and using the integration by parts the actions of the operators $N^{\left(
t\right) }$ and $\gamma $ on $a_{0}^{\mathrm{t-p}}$ are%
\begin{eqnarray}
N^{\left( t\right) }a_{0}^{\mathrm{t-A}} &=&D^{\mu \nu \lambda |\rho }t_{\mu
\nu \lambda |\rho }+\partial _{\mu }n^{\mu },  \label{tpno5} \\
\gamma a_{0}^{\mathrm{t-A}} &=&D^{\mu \nu \lambda |\rho }\gamma t_{\mu \nu
\lambda |\rho }+D^{\mu _{1}...\mu _{p}}\gamma A_{\mu _{1}...\mu
_{p}}+\partial _{\mu }l^{\mu },  \label{tpno6}
\end{eqnarray}%
where we are not interested by the concrete form of the divergences. It
follows from (\ref{tpno6}) that (\ref{tpno1}) has solutions only if the
functional derivatives satisfy the conditions 
\begin{eqnarray}
\partial _{\mu }D^{\mu \nu \lambda |\rho } &=&0,\,\partial _{\rho }D^{\mu
\nu \lambda |\rho }=0,  \label{tpno7} \\
\partial _{\mu }D^{\mu \mu _{2}...\mu _{p}} &=&0.  \label{tpno8}
\end{eqnarray}%
(\ref{tpno7}) implies further using the generalized cohomology of the
space-time exterior differential that $D^{\mu \nu \lambda |\rho }$ must have
the form%
\begin{equation}
D^{\mu \nu \lambda |\rho }=\partial _{\alpha }\partial _{\beta }\Phi ^{\mu
\nu \lambda \alpha |\rho \beta },  \label{tpno9}
\end{equation}%
where the tensor field $\Phi ^{\mu \nu \lambda \alpha |\rho \beta }$ has the
same mixed symmetry as the curvature tensor. We can reconstruct the form of
the solution for the equation (\ref{tpno1}) using (\ref{tpno5}), (\ref{tpno9}%
) and the fact that $a_{0}^{\mathrm{t-A}}$ is eigen solution for $N^{\left(
t\right) }$, $N^{\left( t\right) }a_{0}^{\mathrm{t-A}}=ka_{0}^{\mathrm{t-A}}$%
. Hence, up to a negligible divergence, we obtain a necessary condition for $%
a_{0}^{\mathrm{t-A}}$ to be a solution to (\ref{tpno1}),%
\begin{equation}
a_{0}^{\mathrm{t-A}}=\frac{1}{8k}\Phi ^{\mu \nu \lambda \alpha |\rho \beta
}K_{\mu \nu \lambda \alpha |\rho \beta }.  \label{tpno10}
\end{equation}%
Furthermore we remarque that at the pureghost number zero $\gamma $ splits
in $\gamma =\gamma ^{\left( t\right) }+\gamma ^{\left( A\right) }$ ($\gamma
^{\left( t\right) }$ acts only on the mixed symmetry tensor field and it
derivatives and $\gamma ^{\left( A\right) }$ only on the abelian form and
its derivatives) and in the equation (\ref{tpno1}) every component of $%
\gamma $ must give separately a total derivative. This remarque helps us to
obtain, using almost the same computation technique as in \cite{noijhep31},
that the tensor field $\Phi ^{\mu \nu \lambda \alpha |\rho \beta }$ in (\ref%
{tpno10}) is linear in the fields (it can not depend on the derivatives of
the fields due to the restraint imposed on the derivative order).
Because we study the cross-interactions, we consider that $\Phi ^{\mu \nu
\lambda \alpha |\rho \beta }$ depends only on the abelian form and the
general form of the solution for the equation (\ref{tpno1}) is 
\begin{equation}
a_{0}^{\mathrm{t-A}}=f^{\mu \nu \lambda \alpha |\rho \beta ||\mu _{1}...\mu
_{p}}A_{\mu _{1}...\mu _{p}}K_{\mu \nu \lambda \alpha |\rho \beta },
\label{tpno11}
\end{equation}%
where $f^{\mu \nu \lambda \alpha |\rho \beta ||\mu _{1}...\mu _{p}}$ is a
constant tensor. There is only one solution, for an abelian two-form,%
\begin{equation}
a_{0}^{\mathrm{t-A}}=\sigma _{\lambda \alpha }\sigma _{\rho \beta }A_{\mu
\nu }K^{\mu \nu \lambda \rho |\alpha \beta },  \label{tpno12}
\end{equation}%
but it proves to be $s$-exact modulo $d$,%
\begin{eqnarray}
a_{0}^{\mathrm{t-A}} &=&s\left( \frac{2}{4-D}A_{\mu \nu }t^{\ast \mu \nu }+%
\frac{4}{3\left( 4-D\right) }\sigma _{\lambda \rho }\overset{\left( 1\right) 
}{\xi }_{\nu }\mathcal{G}^{\nu \lambda |\rho }+\frac{2}{3\left( 4-D\right) }%
\overset{\left( 2\right) }{\xi }\mathcal{C}_{\ \ \ \ \nu }^{\ast \nu }\right)
\notag \\
&&+\partial _{\mu }\left( \frac{4}{4-D}\overset{\left( 1\right) }{\xi }_{\nu
}t^{\ast \mu \nu }-\frac{4}{3\left( 4-D\right) }\sigma _{\lambda \rho }%
\overset{\left( 2\right) }{\xi }\mathcal{G}^{\mu \lambda |\rho }\right) .
\label{tpno13}
\end{eqnarray}

\section{Solutions for the first order deformation}

There are four indepent solutions for the first order deformation of the
solution of the master equation. The first two have the nonintegrated
densities given by the formulas (\ref{tpn2})-(\ref{tpn3}) and depend only on
the abelian form (see \cite{citatiicupforme1,citatiicupforme2}),%
\begin{eqnarray}
S_{1}^{\left( 1\right) } &=&\int d^{2p+1}x\epsilon ^{\mu _{1}...\mu
_{2p+1}}A_{\mu _{1}...\mu _{p}}F_{\mu _{p+1}...\mu _{2p+1}},  \label{tpn78}
\\
S_{1}^{\left( 2\right) } &=&\int d^{3p+2}x\epsilon ^{\mu _{1}...\mu
_{3p+2}}A_{\mu _{1}...\mu _{p}}F_{\mu _{p+1}...\mu _{2p+1}}F_{\mu
_{2p+2}...\mu _{3p+2}}.  \label{tpn79}
\end{eqnarray}%
The third solution exist only for an abelian 3-form and it is obtained
mixing two chains of the type (\ref{tp20})-(\ref{tp21}) that begin, both of
them, from the antighost number $I=3$ and with the consistency stoping, for
each separately, at the antighost number one. The first chain has the
components given by the formulas (\ref{tp32})-(\ref{tp34}) and the second by
(\ref{tpn66})-(\ref{tpn68}). If we mix the components (\ref{tp32}) and (\ref%
{tpn66}) with the greatest antighost number,%
\begin{equation}
a_{3}^{^{\prime }}=\mathcal{C}_{\,\,\,\,\,\,\,\,\,\mu }^{\ast \mu }\overset{%
\left( 3\right) }{\xi }+k\overset{\left( 2\right) }{\xi }^{\ast \mu }C_{\mu
},  \label{tpn80}
\end{equation}%
where $k$ is a constant, we get%
\begin{eqnarray}
a_{2}^{^{\prime }\mathrm{t-p}} &=&-2\mathcal{G}^{\ast \lambda \mu }|_{\mu }%
\overset{\left( 2\right) }{\xi }_{\lambda }+k\overset{\left( 2\right) }{\xi }%
^{\ast \nu \mu }C_{\nu \mu },  \label{tpn81} \\
a_{1}^{^{\prime }\mathrm{t-p}} &=&3t^{\ast \rho \lambda \mu }|_{\mu }\overset%
{\left( 1\right) }{\xi }_{\rho \lambda }-\frac{k}{2}A^{\ast \lambda \nu \mu
}\eta _{\lambda \nu \mu },  \label{tpn82}
\end{eqnarray}%
and the two chains are consistent in the zero order together for%
\begin{equation}
k=6\cdot 3!\left( 4-D\right) ,  \label{tpn83}
\end{equation}%
with the next solution of the equation $\delta a_{1}^{^{\prime }\mathrm{t-p}%
}+\gamma a_{0}^{^{\prime }\mathrm{t-p}}=\partial _{\mu }j_{0}^{\mu }$,%
\begin{equation}
a_{0}^{^{\prime }\mathrm{t-p}}=\left( 4-D\right) A^{\rho \lambda \mu
}\partial _{\left[ \theta \right. }t_{\left. \rho \lambda \mu \right]
}|^{\theta }.  \label{tpn84}
\end{equation}%
The formulas (\ref{tpn80})-(\ref{tpn84}) give us as the components of a
solution for the first order deformation in the case of the
cross-interactions for an abelian 3-form and the tensor field $t_{\lambda
\mu \nu |\rho }$,%
\begin{eqnarray}
S_{1}^{\left( 3\right) } &=&\int d^{D}x\left( \mathcal{C}_{\,\,\,\,\,\,\,\,%
\,\mu }^{\ast \mu }\overset{\left( 3\right) }{\xi }+k\overset{\left(
2\right) }{\xi }^{\ast \mu }C_{\mu }-2\mathcal{G}^{\ast \lambda \mu }|_{\mu }%
\overset{\left( 2\right) }{\xi }_{\lambda }+k\overset{\left( 1\right) }{\xi }%
^{\ast \nu \mu }C_{\nu \mu }+\right.   \notag \\
&&\left. 3t^{\ast \rho \lambda \mu }|_{\mu }\overset{\left( 1\right) }{\xi }%
_{\rho \lambda }-\frac{k}{2}A^{\ast \lambda \nu \mu }\eta _{\lambda \nu \mu
}+\left( 4-D\right) A^{\rho \lambda \mu }\partial _{\left[ \theta \right.
}t_{\left. \rho \lambda \mu \right] }|^{\theta }\right) .  \label{tpn85}
\end{eqnarray}%
The fourth independent solution for the first order deformation has the
components written in the formulas (\ref{tp68})-(\ref{tp69}),%
\begin{eqnarray}
S_{1}^{\left( 4\right) } &=&\int d^{p+4}x\ \epsilon ^{\mu _{1}...\mu
_{p+4}}\left( A_{\mu _{1}...\mu _{p}}^{\ast }\partial _{\lbrack \mu
_{p+1}}\eta _{\mu _{p+2}\mu _{p+3}\mu _{p+4}]}+\right.   \notag \\
&&\left. \frac{1}{3p!}F_{\lambda \mu _{1}...\mu _{p}}\partial _{\lbrack \mu
_{p+1}}t_{\mu _{p+2}\mu _{p+3}\mu _{p+4}]}|^{\lambda }\right) .
\label{tpn86}
\end{eqnarray}%
We remark, first, the solutions $S_{1}^{\left( 1\right) }$ and $%
S_{1}^{\left( 2\right) }$ can not appear together, because of the space-time
dimensions incompatibility ($2p+1\neq 3p+2,\ \forall p\geq 1$). Second, $%
S_{1}^{\left( 1\right) }$ and $S_{1}^{\left( 4\right) }$ can appear together
only if $2p+1=p+4$ ($p=3,\ D=7$), while $S_{1}^{\left( 2\right) }$ and $%
S_{1}^{\left( 4\right) }$ can appear together only if $3p+2=p+4$ ($p=1,\ D=5$%
; this case was analysed in \cite{31vect}). To conclude, for an abelian
3-form the general solution for the first order deformation is%
\begin{equation}
S_{1}=\delta _{D,7}\left( c_{1}S_{1}^{\left( 1\right) }+c_{4}S_{1}^{\left(
4\right) }\right) +\delta _{D,11}c_{2}S_{1}^{\left( 2\right)
}+c_{3}S_{1}^{\left( 3\right) },  \label{tpn87}
\end{equation}%
where $c_{1},\ c_{2},\ c_{3},\ c_{4}$ are constants, while for an abelian
form with $p\notin \left\{ 1,3\right\} $ the general solution is only $%
S_{1}^{\left( 4\right) }$.

\section{Higher order deformations}

Second order deformation $S_{2}$ is the solution of the equation (\ref{tpn02}%
). The first order deformation for the case of the cross-interactions
between an abelian $p$-form and the mixed symmetry $\left( 3,1\right) $
tensor field has the general form (\ref{tpn86}) for every value $p\geq 1$ of
the form degree. For the particular case of a three-form ($p=3$) the first
order deformation is given by (\ref{tpn87}), but in our estimations it is
consistent (i.e. the equation (\ref{tpn02}) has a solution) only if the
coefficients $c_{1}$, $c_{2}$, $c_{3}$ vanish. So, we will take into account
only the solution $S_{1}^{\left( 4\right) }$ given by (\ref{tpn86}) for the
first order deformation and the first term in (\ref{tpn86}) is%
\begin{eqnarray}
\frac{1}{2}\left( S_{1},S_{1}\right)  &=&\frac{1}{2}\int d^{D}x\ \frac{1}{3p}%
\epsilon ^{\nu _{1}...\nu _{p}\mu _{p+1}...\mu _{p+4}}\epsilon _{\nu
_{1}...\nu _{p}\nu _{p+1}...\nu _{p+4}}\times   \notag \\
&&\partial _{\lambda }\partial _{\lbrack \mu _{p+1}}\eta _{\mu _{p+2}\mu
_{p+3}\mu _{p+4}]}\partial ^{\lbrack \nu _{p+1}}t^{\nu _{p+2}\nu _{p+3}\nu
_{p+4}]|\lambda },  \label{tp71}
\end{eqnarray}%
from which using the identity%
\begin{equation}
\epsilon ^{\nu _{1}...\nu _{p}\mu _{p+1}...\mu _{p+4}}\epsilon _{\nu
_{1}...\nu _{p}\nu _{p+1}...\nu _{p+4}}=\left( -1\right) ^{p+1}\left(
p!\right) \delta _{\lbrack \nu _{p+1}}^{\mu _{p+1}}...\delta _{\nu
_{p+4}]}^{\mu _{p+4}}  \label{tp72}
\end{equation}%
it follows 
\begin{equation}
\frac{1}{2}\left( S_{1},S_{1}\right) =\frac{1}{2}\int d^{D}x\ s\left( \left(
-1\right) ^{p+1}\frac{32}{3}\partial ^{\lbrack \lambda }t_{\ \ \ \ \ \ \
|\lambda }^{\mu \nu \rho ]}\partial _{\lbrack \alpha }t_{\mu \nu \rho ]}^{\
\ \ \ \ \ \ |\alpha }\right) .  \label{tp73}
\end{equation}%
Therefore we have for the second order deformation the solution 
\begin{equation}
S_{2}=\int d^{D}x\ \left( \left( -1\right) ^{p}\frac{16}{3}\partial
^{\lbrack \lambda }t_{\ \ \ \ \ \ \ |\lambda }^{\mu \nu \rho ]}\partial
_{\lbrack \alpha }t_{\mu \nu \rho ]}^{\ \ \ \ \ \ \ |\alpha }\right) .
\label{tp74}
\end{equation}%
Because $\left( S_{1},S_{2}\right) =0$ we can choose the deformations of the
order higher than two to vanish,%
\begin{equation}
S_{k}=0,\ k\geq 3.  \label{tp75}
\end{equation}

\section{Conclusions}

The deformation of the solution of the master equation for the theory
described by the action (\ref{tp1}) is consistent in a space-time with the
dimension $D=p+4$, if the development according to the antighost number of
the first order deformation has only two terms (\ref{tp67}), that with the
antighost number one being (\ref{tp68}). The components of the deformation $%
\bar{S}=S+gS_{1}+g^{2}S_{2}$ are written $S$ in (\ref{tp15}), $S_{1}$ in (%
\ref{tp70}) and $S_{2}$ in (\ref{tp74}). The terms with the antighost number
one from $\bar{S}$ represents the deformed gauge transformations. Thus, we
observe that the gauge transformations of the tensor field $t_{\lambda \mu
\nu |\alpha }$ are unchanged, while the gauge transformations of the abelian
form are%
\begin{equation}
\delta _{\overset{\left( 1\right) }{\rho },\ \epsilon }=\partial _{\left[
\mu _{1}\right. }\overset{\left( 1\right) }{\rho }_{\left. \mu _{2}...\mu
_{p}\right] }+g\epsilon ^{\mu _{1}...\mu _{p}\lambda \mu \nu \rho }A_{\mu
_{1}...\mu _{p}}^{\ast }\partial _{\left[ \lambda \right. }\epsilon _{\left.
\mu \nu \rho \right] },  \label{tp101}
\end{equation}%
where $\epsilon _{\mu \nu \rho }$ are the gauge parameters from (\ref{tp4}).
The terms with the antighost number zero from $\bar{S}$ represent the
Lagrangian action of the interacting theory, which can be written as%
\begin{equation}
S_{L}\left[ t_{\mu \nu \lambda |\rho },A_{\mu _{1}...\mu _{p}}\right] =S_{0}%
\left[ t_{\mu \nu \lambda |\rho }\right] -\frac{1}{2\left( p+1\right) !}\int
d^{p+4}x\ \bar{F}_{\mu _{1}\ldots \mu _{p+1}}\bar{F}^{\mu _{1}\ldots \mu
_{p+1}},  \label{tp102}
\end{equation}%
where $\bar{F}_{\mu _{1}\ldots \mu _{p+1}}$ are the most general objects
depending on the abelian form, invariant under the gauge transformations (%
\ref{tp4}) and (\ref{tp101}),%
\begin{equation}
\bar{F}_{\mu _{1}\ldots \mu _{p+1}}=F_{\mu _{1}\ldots \mu _{p+1}}-\frac{g}{3}%
\partial ^{\left[ \lambda \right. }t_{\ \ \ \ \ \ \ \ |\left[ \mu
_{1}\right. }^{\left. \mu \nu \rho \right] }\epsilon _{\left. \mu _{2}\ldots
\mu _{p+1}\right] \lambda \mu \nu \rho }.  \label{tp103}
\end{equation}%
($F_{\mu _{1}\ldots \mu _{p+1}}$ is defined in (\ref{tp3}).) The terms which
contain products of the abelian form with the mixed symmetry $\left(
3,1\right) $ tensor field does not represent interaction vertices, being
known in the literature as \textquotedblleft mixing terms\textquotedblright .

\end{document}